\documentclass[letterpaper,11pt]{article}
\pdfoutput=1 
\usepackage{caption}
\usepackage{xcolor}
\usepackage{comment}
\usepackage{enumerate}
\usepackage{framed}
\usepackage{mdframed}
\usepackage{amsmath}
\usepackage{amsfonts}
\usepackage{mathrsfs}
\usepackage{euscript}
\usepackage{amssymb}
\usepackage[normalem]{ulem}
\usepackage{enumitem}  
\usepackage{graphicx, rotating}
\usepackage{epsfig}
\usepackage{latexsym}
\usepackage{graphicx}
\usepackage{color}
\usepackage{xcolor}
\usepackage{amsmath,bm,amssymb}
\usepackage{cite}
\usepackage{slashed}
\usepackage{hyperref}
\usepackage{epstopdf}
\usepackage[title]{appendix}
\usepackage[font=footnotesize,labelfont=]{caption}
\AppendGraphicsExtensions{.tif}
\hypersetup{colorlinks=true, citecolor=bluscuro, linkcolor=black, urlcolor=bluscuro}
\definecolor{rosso}{cmyk}{0,1,1,0.55}
\definecolor{bluscuro}{rgb}{0.15, 0.2, .85}
\definecolor{bluchiaro}{cmyk}{1,.3,0.,0.1}

\newcommand{\be}{\begin{equation}}
\newcommand{\ee}{\end{equation}}
\newcommand{\bea}{\begin{eqnarray}}
\newcommand{\eea}{\end{eqnarray}}
\newcommand{\beas}{\begin{eqnarray*}}
\newcommand{\eeas}{\end{eqnarray*}}

\newcommand{\rd}{{\rm d}}
\newcommand{\ri}{{\rm i}}

\def\beq{\begin{eqnarray}}
\def\eeq{\end{eqnarray}}
\newcommand{\la}{\langle}
\newcommand{\ra}{\rangle}

\begin{document}
\def\thefootnote{\fnsymbol{footnote}}

\begin{center}
\LARGE{\textbf{Coherence and Quantum Stability of Relativistic Superfluid States}  \\[0.5cm]

\large{{Lasha Berezhiani,$^{1,2}$\footnote{lashaber@mpp.mpg.de} Giordano Cintia,$^{3,4}$\footnote{giordano.cintia@univ-amu.fr} Giacomo Contri.$^{1,2}$\footnote{contri@mpp.mpg.de}
}}\\[0.5cm]

\small{
\textit{
$^1$Max-Planck-Institut f\"ur Physik, Werner–Heisenberg–Institut, \\ Boltzmannstra{\ss}e~8,
85748 Garching, Germany\\
 \vskip 5pt
~~$^2$ Arnold Sommerfeld Center, Ludwig-Maximilians-Universit\"at, \\Theresienstra{\ss}e 37, 80333 M\"unchen, Germany
\\
 \vskip 5pt
~~$^3$ Galileo Galilei Institute for Theoretical Physics, Largo Enrico Fermi, 2, 50125 Firenze, Italy
\\
 \vskip 5pt
~~$^4$ Aix Marseille Universit\'e, Universit\'e de Toulon, CNRS, CPT, Marseille, France}
 }

 \vspace{.2cm}

}
\end{center}

\vspace{.6cm}

\hrule \vspace{0.2cm}
\centerline{\small{\bf Abstract}}
\vspace{0.1cm}
{\small\noindent 
We analyze the quantum dynamics of a relativistic homogeneous superfluid in a complex scalar field theory. Unlike zero-charge condensates, which undergo quantum evaporation due to internal number-changing processes, we show that $U(1)$ superfluids preserve their internal coherence indefinitely in this theory.
In particular, although not Hamiltonian eigenstates, these configurations are stable in the full quantum theory to all orders in perturbation theory. This is demonstrated by explicitly constructing the corresponding quantum state and  
studying its dynamics.
Crucially, maintaining stability requires the quantum state to go beyond a naive coherent-state construction: specific non-Gaussian corrections are essential for having a stationary state. The resulting state is identified as the interacting vacuum of the superfluid fluctuations, which also serves as the ground state of the modified Hamiltonian $\hat{{H}}-\mu \hat{Q}$, with $\mu$ the full-fledged quantum chemical potential and $\hat{Q}$ the $U(1)$ charge.
Finally, we check that the phonon mode remains gapless once one-loop corrections are included, confirming the robustness of the Goldstone theorem beyond the semiclassical regime, even in systems with a spontaneously broken Lorentz symmetry.
}

\vspace{0.3cm}
\noindent
\hrule
\def\thefootnote{\arabic{footnote}}
\setcounter{footnote}{0}
\newpage
\tableofcontents
\newpage
\section{Introduction and summary}

Coherent states provide an important connection between classical and quantum theories~\cite{Glauber:1963tx, Sudarshan:1963ts,Kibble:1965zza}\footnote{See also \cite{Zhang:1990fy,Zhang:1999is}.}. However, their self-consistent construction is often challenging in interacting theories,
since there is no general procedure to build them directly from
the fundamental degrees of freedom. 
This problem is often circumvented by means of the semiclassical approximation, in which the fundamental quantum field is expanded around a classical configuration that defines the new vacuum of the theory, and the fluctuations around this background are then quantized to define a new Hilbert space. 

While this is a powerful tool for quantifying certain quantum corrections to these configurations, it comes with some limitations.
From a fundamental standpoint, the theory must first be quantized around a well-defined vacuum, which is the lowest energy eigenstate of the Hamiltonian, determine its Hilbert space, and then identify the class of states from which the semiclassical approach emerges naturally. 
The importance of this approach is most pronounced for gravitational systems \cite{Dvali:2011aa, Dvali:2012en, Dvali:2013eja, Dvali:2014gua, Berezhiani:2016grw, Dvali:2017eba, Berezhiani:2021zst,Berezhiani:2024boz}, for which the Minkowski spacetime is the only one that can serve as a well-defined $S$-matrix vacuum. See also \cite{Dvali:2013vxa,Dvali:2017ruz,Berezhiani:2020pbv,Berezhiani:2021gph,Dvali:2022vzz,Berezhiani:2023uwt} for similar discussions in non-gravitational quantum field theories. 

One of the essential points is that the promotion of classical configurations to vacua is not always reliable, as not all stationary classical configurations remain stable at the quantum level. Interactions among the quantum constituents of the system are expected to invalidate the classical description due to the quantum loss of coherence. For physically interesting systems, the question is how the time-scale of this effect, coined as \textit{quantum break time} in \cite{Dvali:2013vxa}, compares to relevant classical time-scales. 
The importance of such effects has been identified in several systems of cosmological interest such as black holes \cite{Dvali:2011aa, Dvali:2012en, Dvali:2013eja}, cosmic inflation \cite{Dvali:2013eja, Berezhiani:2016grw, Berezhiani:2015ola, Berezhiani:2022gnv} and de Sitter spacetime \cite{Dvali:2013eja, Dvali:2017eba}.\footnote{An example of an effect not captured by the semiclassical description is the so-called memory burden effect\cite{Dvali:2018xpy,Dvali:2020wft}. The idea is that if a macroscopic quantum state can store information in a large number of nearly gapless microscopic modes, then the memory burden may provide a stabilizing role in systems that are (semi-)classically unstable. Recently, this effect has been proposed as a mechanism that may slow down the evaporation of black holes after their half-life, thereby opening a new mass window for primordial black holes as a dark matter candidate below $10^{15}{\rm g}$ \cite{Alexandre:2024nuo, Dvali:2024hsb, Dvali:2025ktz}.}

The simplest example of this class of effects is provided by the dynamics of a condensate of interacting real scalar bosons.
By analyzing the evolution of a homogeneous coherent state in a theory with quartic interactions, it has been shown in \cite{Baacke:1996se, Dvali:2017ruz, Berezhiani:2021gph}
that these configurations 
 are depleted by number-changing processes. In particular, the annihilation of zero-momentum constituents into relativistic ones leads to the damping of the one-point function of the field operator on the coherent state. 
 A key feature is that, even if the real condensate is prepared to be maximally classical (by initializing it as a coherent state that saturates the uncertainty principle) interactions among internal constituents inevitably enhance the quantumness of the system, ultimately invalidating the semiclassical approximation.

 In this context, it might seem to be a generic feature that finite density 
 states of interacting degrees of freedom should undergo quantum breaking. Since this class of states does not correspond to Hamiltonian eigenstates, it is natural to expect their time evolution to be nontrivial, and that any classically initialized configuration becomes increasingly more quantum as time progresses. 
 
This, however, can be prevented by underlying symmetries.
In this work, we will show that for a complex scalar field the presence of a conserved $U(1)$ charge allows for protected states. In particular, we will study the quantum state, with coherent features, which corresponds to the homogeneous superfluid of massive self-interacting bosons. As we shall see, these protected configurations exhibit a stationary time evolution despite not being eigenstates of the Hamiltonian. 
Moreover, we will emphasize that a correct initialization of higher-order correlation functions (beyond the one-point function that fixes the initial background configuration) is essential for ensuring the stability of the background once quantum effects are taken into account. This follows from the fact that, in interacting theories, correlation functions satisfy an infinite hierarchy of coupled equations of motion. Consequently, the evolution of the background cannot be consistently decoupled from the initial conditions imposed on the connected part of higher-order correlators.


Investigating the quantum stability of interacting field configurations has several important consequences. Although an infinitely extended homogeneous condensate is clearly unrealistic, as it would carry infinite total energy (despite having finite energy density) and cannot be realized experimentally, it nevertheless provides a useful idealization in cosmological scenarios. 
\\
Generically, in the fundamental quantum description, cosmological backgrounds must be represented by a quantum state, with a certain degree of coherence. These states cannot be eigenstates of the Hamiltonian due to their non-trivial time-evolution and/or semiclassical particle production. Coherent states, when considered with appropriate dressings as discussed in ref~\cite{Berezhiani:2023uwt} and expanded below for superfluid configurations, are expected
to capture essential aspects of these systems. For (semi-)classically (quasi-)eternal backgrounds, it is crucial to test this apparent eternity within a fundamental quantum framework, for instance, through the coherent-state construction.\footnote{For example, the simplest model of dark matter superfluidity is based precisely on a coherent state of massive repulsively interacting particles \cite{Goodman:2000tg, Slepian:2011ev, Berezhiani:2015pia, Berezhiani:2015bqa, Berezhiani:2021rjs, Berezhiani:2022buv, Berezhiani:2018oxf} (see also \cite{Berezhiani:2025maf} for a recent review). For the real scalar field case the parameter space is restricted by quantum depletion of the condensate \cite{Berezhiani:2022buv}. However, it has been argued that the number-changing depletion channels of the real field could be suppressed by considering a complex field, due to charge conservation. This naturally leads to the analysis of configurations of the type considered in this paper.}

Before presenting our main results, we briefly discuss, in the remainder of this introduction, the relation between classical and quantum configurations in general terms.

\subsection{States and classicality}
In classical field theory, the state of the system is described by c-number field configurations, which we denote collectively as
${\Phi}(x)$ for the sake of a general argument. In quantum field theory, by contrast, the system is described by the quantum state $|\Psi\ra$. This is equivalent to specifying the infinite set of correlation functions of the underlying quantum operators, collectively denoted by $\hat{\phi}(x)$.  
Consequently, while in the classical theory the initial value problem is posed in terms of the initial value of $\Phi$ and its time-derivative, the quantum state in principle provides infinitely more information in terms of correlation functions, e.g.
\beq
&&\la \Psi | \hat{\phi}(x) |\Psi\ra\,,\nonumber\\
&&\la \Psi | \hat{\phi}(x)\hat{\phi}(y) |\Psi\ra\,,\nonumber\\
&&\la \Psi | \hat{\phi}(x) \hat{\phi}(y)\hat{\phi}(z)|\Psi\ra\,,\nonumber\\
&&\ldots\nonumber
\eeq
and similar correlators involving the conjugate momenta.

Classicality of the system implies that its quantum constituent degrees of freedom are in a state of high occupancy. 
This allows us to define the classical field configuration as the one-point function of a relevant operator\footnote{It must be stressed that the latter, in general, is not necessarily related to the expectation value. Instead, it could be a different matrix element of the field operator. Although this is an interesting and well-understood point \cite{Dvali:2013eja}, we will carry on with our general discussion under the assumption that the classicality is encoded in the expectation value. This is the case, e.g., for coherent states, which are of our primary interest in this work.} $\la \Psi | \hat{\phi}(x) |\Psi\ra\equiv \Phi(x)$, and guarantees that the one-point function alone provides an adequate description of
the system. This entails that, on the relevant scales of the problem at hand, higher-order correlation functions are dominated by their disconnected contributions, i.e.

\beq
&&\la \Psi | \hat{\phi}(x)\hat{\phi}(y) |\Psi\ra\simeq \la \Psi | \hat{\phi}(x)|\Psi\ra \la\Psi|\hat{\phi}(y) |\Psi\ra\,,\nonumber\\
&&\la \Psi | \hat{\phi}(x) \hat{\phi}(y)\hat{\phi}(z)|\Psi\ra\simeq \la \Psi | \hat{\phi}(x)|\Psi\ra\la\Psi| \hat{\phi}(y)|\Psi\ra\la\Psi|\hat{\phi}(z)|\Psi\ra\,,\nonumber\\
&&\ldots\,.\nonumber
\eeq

One should keep in mind that the scope of such statements is scale-dependent. At sufficiently short distances, the connected parts of correlation functions always dominate, indicating the unavoidable breakdown of classicality as quantum fluctuations will dominate over the background.
Furthermore, even if the system is initially prepared in a state with such classical features, it is by no means guaranteed to withstand the passage of time. 
The persistence of classicality is connected to the validity of the above relations under the Hamiltonian flow.

An excellent example of a classical state is a coherent state. It is straightforward to construct it in non-interacting theories, or alternatively as asymptotic states of the interacting theory, by using the field displacement operator on the vacuum. In the Schrödinger picture, and focusing on a scalar theory, it is built as
\beq
|\Psi\ra=e^{-\ri\int \rd^3 x \left(\phi_{\rm cl}(\vec{x})\hat{\Pi}(\vec{x})-\Pi_{\rm cl}(\vec{x})\hat{\phi}(\vec{x})\right)}|0\ra\,.
\eeq
By construction, this state sets the initial expectation values of the canonical variables to $\phi_{\rm cl}(\vec{x})$ and $\Pi_{\rm cl}(\vec{x})$, respectively. In a free theory, it is easy to show that the evolution of this state results in replacing these classical functions by the time-dependent solution to the equation of motion with corresponding initial conditions. Hence, the classicality of this state is preserved.

The moment we make the theory interacting,  subtleties begin to emerge \cite{Berezhiani:2020pbv, Berezhiani:2021gph, Berezhiani:2023uwt}. For starters, one would be inclined to replace the free vacuum with the
vacuum of the interacting theory, providing 
\beq
|\Psi\ra=e^{-{\rm i}\int {\rm d}^3 x \left( \phi_{\rm cl}(\vec{x})\hat{\Pi}(\vec{x})-\Pi_{\rm cl}(\vec{x})\hat{\phi}(\vec{x}) \right)}|\Omega\ra\,.
\label{intro:coherent}
\eeq
This state would set the following initial conditions for the one-point functions of the canonical variables:
\beq
\label{initphi}
&&\la \Psi | \hat{\phi} |\Psi\ra (t=0)=\phi_{\rm cl}(\vec{x})\\
&&\la \Psi | \hat{\Pi} |\Psi\ra (t=0)=\Pi_{\rm cl}(\vec{x})\,,
\label{initpi}
\eeq
together with the following two-point functions
\beq
\label{init2phi}
&&\la \Psi | \hat{\phi}(\vec{x},0)\hat{\phi}(\vec{y},0) |\Psi\ra=\phi_{\rm cl}(\vec{x})\phi_{\rm cl}(\vec{y})+\la \Omega | \hat{\phi}(\vec{x},0)\hat{\phi}(\vec{y},0) |\Omega\ra\,\\
&&\la \Psi | \hat{\Pi}(\vec{x},0)\hat{\Pi}(\vec{y},0) |\Psi\ra=\Pi_{\rm cl}(\vec{x})\Pi_{\rm cl}(\vec{y})+\la \Omega | \hat{\Pi}(\vec{x},0)\hat{\Pi}(\vec{y},0) |\Omega\ra\,,
\label{init2pi}
\eeq
and similarly for higher-order correlators.
However, the appearance of the vacuum two-point function in the above expression leads to a non-renormalizable infinite energy density for the configuration under consideration \cite{Baacke:1997zz, Berezhiani:2021gph}. To fix this, the vacuum of the theory $|\Omega\ra$ 
must undergo a background-dependent squeezing to ensure one-loop consistency, and even non-Gaussian modifications at higher loops \cite{Berezhiani:2023uwt}. 

\subsection{Time-evolution and the background field method}
Once we have a consistent non-asymptotic construction of a state, we could study its departure from classicality in various ways. The most obvious one involves its direct time evolution, that in Schrödinger picture corresponds to
\beq
|\Psi\ra(t)=e^{-{\rm i}\hat{H}t}e^{-{\rm i}\int {\rm d}^3 x \left( \phi_{\rm cl}(\vec{x},0)\hat{\pi}(\vec{x})-\pi_{\rm cl}(\vec{x},0)\hat{\phi}(\vec{x}) \right)}|\Omega\ra
\label{intro:quantumevolution}\,.
\eeq
The departure from classicality can then be assessed by comparing it with the classical state
\beq
|\Psi_{\rm cl}\ra(t)=e^{-{\rm i}\int {\rm d}^3 x \left( \phi_{\rm cl}(\vec{x},t)\hat{\pi}(\vec{x})-\pi_{\rm cl}(\vec{x},t)\hat{\phi}(\vec{x}) \right)}|\Omega\ra\,,
\eeq
which assumes that the entire dynamics is confined within the time-evolution of the c-number functions of \eqref{intro:coherent}, determined by the classical equations of motion. The departure can be quantified by the overlap of this state with the quantum one, which could in principle be organized as a time series 
\beq
|\la \Psi_{\rm cl}| \Psi\ra|^2(t)=1-\sum_{n} \alpha_n t^n\,.
\eeq
The quantum break time emerges as the time scale at which the time-dependent part becomes significant.

However, this approach can be challenging to implement for physically interesting cases. Instead, for computational practicality, it is sometimes convenient to think about the state in terms of the correlation functions. As we have already pointed out, the two descriptions are equivalent. 
As a result, rather than using \eqref{intro:quantumevolution}, we can study the coupled, infinite-dimensional system of equations governing the correlation functions.\footnote{These are usually called Schwinger–Dyson equations. In contrast to the standard derivation, however, in this work we formulate them as equations of motion for non-time-ordered correlation functions.} 
This approach becomes considerably more practical than solving the full quantum evolution of the state when there exists a certain hierarchy among the connected and disconnected parts of correlation functions. In such cases, at a given order of the loop expansion, the system of equations simplifies so that only a finite number of correlators mutually source each other.

The derivation of this system of equations is best demonstrated employing the background field method for quantum states with a classical one-point function, which, incidentally, is precisely the type of states we are interested in here. For illustrative purposes, we will use the case of the real scalar field with quartic interactions. The specific details of the theory are irrelevant, as the mere presence of nonlinearities is enough to illustrate the qualitative behavior. 
In the Heisenberg picture, the full quantum dynamics is governed by Hamilton's equations, which for the quartic theory under consideration reduce to the following second-order differential equation
\beq
\left( \Box+m^2\right)\hat{\phi}(x)+\frac{\lambda}{3!}\hat{\phi}^3(x)=0\,.
\label{intro:equation}
\eeq
Next, we define the background field as the fully quantum one-point function $\Phi\equiv\la\Psi|\hat{\phi}(x)|\Psi\ra$ rather than just the classical background, as well as the operator $\hat{\psi}\equiv\hat{\phi}-\Phi$ characterizing the deviation from the background, which has a vanishing expectation value in state $|\Psi\ra$ by definition. Substituting this decomposition back in \eqref{intro:equation} and taking the expectation value, we arrive at
\beq
\label{intro:bckgreq}
\left( \Box+m^2 +\frac{\lambda}{2}\la \Psi | \hat{\psi}^2(x) |\Psi\ra \right)\Phi(x)+\frac{\lambda}{3!}\,\Phi^3(x)+\frac{\lambda}{3!}\la \Psi | \hat{\psi}^3(x) | \Psi\ra=0\,.
\eeq
This is the quantum extension of the classical equation of motion, with quantum corrections to the background sourced by higher-order correlation functions, which in turn satisfy analogous equations that couple them to even higher-order correlation functions. If we are interested in leading order quantum corrections to the background dynamics we shall ignore the last term of \eqref{intro:bckgreq}, which induces two-loop corrections. In other words, capturing one-loop effects in the above equation requires merely the knowledge of the two-point function. This, in turn, is governed by its own equation of motion\footnote{See Ref.~\cite{Berezhiani:2021gph} for the derivation of the above equation. It is obtained by plugging the decomposition $\hat{\psi}\equiv\hat{\phi}-\Phi$ into \eqref{intro:equation}, subtracting the equation of motion for the one-point function, multiplying by a power of $\psi(x')$, and finally considering the expectation value over $|\Psi\rangle$.}
\beq
\label{2pointpsi}
&&0=\left( \Box^{(x)}+m^2 +\frac{\lambda}{2}\,\Phi^2(x) \right)\la \Psi|\hat{\psi}(x)\hat{\psi}(x')|\Psi\ra\nonumber\\
&&~~~~~~+\frac{\lambda}{2}\,\Phi(x) \la \Psi | \hat{\psi}^2(x)\hat{\psi}(x') | \Psi\ra+\frac{\lambda}{3!}\la \Psi | \hat{\psi}^3(x)\hat{\psi}(x') | \Psi\ra\,.
\eeq
Similar equations for all correlation functions straightforwardly follow from \eqref{intro:equation}. 

The incorporation of quantum corrections can be carried out iteratively in orders of $\hbar$. 
For example, to find the one-loop corrected $\Phi$ from~\eqref{intro:bckgreq}, we shall solve \eqref{2pointpsi} at the tree level. This corresponds to replacing $\Phi$ in the first line by the classical background, and dropping the second line. 
To obtain the one-loop corrected two-point function instead, we need to use the one-loop corrected background field $\Phi$ in the first line of \eqref{2pointpsi}. However, now the second line gives honest one-loop corrections as well, descending from tree-level higher order correlators. 
It is important to keep in mind that these contributions cannot be reduced to merely the correction of $\Phi$.

To recap, the initial quantum state of the system sets the initial conditions for all correlation functions, which then evolve according to the coupled system of equations \eqref{intro:bckgreq}, \eqref{2pointpsi}, and their higher-order analogues. 
Furthermore, as the construction suggests, the background field method can be used to specify the state using the fluctuation field $\hat{\psi}$ order-by-order in perturbation theory and piecewise in terms of its correlation functions. Since fluctuations live on the time-dependent background, in general the effective Hamiltonian for $\hat{\psi}$ would be explicitly time dependent. As such, we would not have the notion of an absolute vacuum for fluctuations, and one of the possibilities would be to consider the instantaneous vacuum as an initial state. However, in reality, we should keep in mind that it all comes down to what initial state of the system we are interested in. 

Either way, the consistent way to think about this procedure is not the quantization of perturbations around the classical background. Instead, one identifies a nontrivial state within the already quantized theory and rewrites the theory in terms of the fluctuation operator $\hat{\psi}$, followed by a convenient mode expansion for it. This provides an alternative identification procedure for the original state. Most importantly, the practical convenience of the identification of the state, either in terms of the original degree of freedom $\hat{\Phi}$ or the fluctuation $\hat{\psi}$, depends on the problem at hand.

Before moving to the analysis of the main system considered in this paper, we conclude the introduction by highlighting the advantages of our framework in clarifying several commonly misunderstood aspects of black hole evaporation. In particular, consider Equations \eqref{intro:bckgreq} and \eqref{2pointpsi}.
In the low-energy effective theory on Minkowski vacuum, the classical black hole geometry plays the role of the background field $\Phi$, while the Hawking radiation appears in the two-point function of metric fluctuations, given by the analog of \eqref{2pointpsi}. The evaporation spectrum is not fixed by the background metric alone; it also requires the second line of \eqref{2pointpsi}, which encodes fluctuation interactions. Moreover, as the black hole mass decreases, the background becomes time-dependent, making the mode-function analysis a genuinely time-dependent problem. In fact, as it has been argued in \cite{Dvali:2015aja}, merely this quantum time-dependence of the background modifies the particle-production spectrum so much as to exclude a self-similar thermal evaporation, and provides corrections large enough to nullify the so-called black hole information paradox.

The structure of the paper is as follows. In Section \ref{sec:state_def}, we give the definition of the stable superfluid state, and comment on its basic properties. In Section \ref{sec3}, we first discuss the semiclassical and quantum description of a superfluid by means of the background field method. Then, we explicitly formulate the definition of the superfluid state above in that language, and we prove its stability at all orders. 
Finally, we expand on the interpretation of the stable superfluid state as a Spontaneous Symmetry Probing state, and on the role of the symmetry in ensuring stability. In Section \ref{sec:stability}, we begin by deriving the one-loop corrections to the relativistic chemical potential, and we show their finiteness after renormalization. Then, we proceed to explain how stability is realized in perturbation theory. We show in the specific counterexample of a Gaussian superfluid state that modifications of the original state result in a non-trivial departure from classical dynamics, at two-loop order. Finally, in Section \ref{sec:gaprel}, we study the spectrum of the fluctuations around the superfluid at one-loop. In particular, we explicitly demonstrate that the phonon remains massless when leading quantum corrections are included. In Section \ref{sec:outlook}, we conclude and provide an outlook for this work.

\textbf{Conventions:} We adopt the mostly-minus signature for the spacetime metric and set $\hbar=c=1$. In the main body of the paper, we drop hats on operators.

\section{A stable superfluid state}\label{sec:state_def}
In this work, we study the stabilization of the homogeneous condensate of an interacting scalar field, carrying the global $U(1)$ charge, which we will be referring to as a superfluid state. Based on the above discussion, a coherent state (with certain squeezing adjustments for consistency) is the prime candidate for the fundamental description of such a system. 

Coherent states are not eigenstates of the Hamiltonian; consequently, they evolve in time. While in a free theory coherence is maintained, in the presence of nonlinearities this is not expected in general to be the case. Nevertheless, we will demonstrate the aforementioned homogeneous superfluid state to retain its full coherence perturbatively to all orders in the quantum coupling, upon an appropriate definition of its state.

We consider a complex scalar field with quartic interactions, governed by the Lagrangian density
\begin{equation}
\mathcal{L} = \partial_\mu \Phi^\dagger \partial^\mu \Phi - m^2 |\Phi|^2 - \lambda |\Phi|^4\,.
\label{eq:LagrSF}
\end{equation}
The classical equation of motion of this theory possesses a stable homogeneous superfluid solution for positive $\lambda$ and $m^2$, parameterized by a single free parameter and given by
\beq
\Phi=\frac{1}{\sqrt{2}}ve^{\ri\mu t}\,,\qquad \text{with} \qquad \mu^2=m^2+\lambda v^2\,.
\label{clsuperfluid}
\eeq
This configuration breaks the global $U(1)$ symmetry, responsible for the particle number conservation, spontaneously. It also breaks some of the spacetime symmetries, due to the explicit time-dependence, nonvanishing energy and $U(1)$ charge density. The latter is given by $n=\mu v^2$. We can choose either one of $n,v,{\rm or}~\mu$ as a free parameter characterizing the superfluid configurations.

In order to identify the quantum state associated with this classical superfluid configuration \eqref{clsuperfluid}, we proceed with a canonical quantization of the theory given by \eqref{eq:LagrSF}. The Hamiltonian is given by
\beq
\hat{H}=\int \rd^3x\left( |\hat{\Pi}|^2 +|\vec{\nabla}\hat{\Phi}|^2+m^2|\hat{\Phi}|^2+\lambda |\hat{\Phi}|^4\right)\,.
\label{H4}
\eeq
We can then proceed with constructing the state corresponding to the background \eqref{clsuperfluid}, say at $t=0$, similar to the analysis of the real field performed in \cite{Berezhiani:2021gph}. In particular, we can consider the state
\beq
|v\ra=e^{-\frac{\ri v}{\sqrt{2}}\int \rd^3x\left(\hat{\Pi}+\ri\mu\hat{\Phi}+\text{h.c.}\right)}|\tilde{\Omega}\ra\,,
\label{intro:coherentsupefluid}
\eeq
with $|\tilde{\Omega}\ra$ denoting a state in which canonical degrees of freedom have vanishing expectation values.
In general, this state cannot be chosen to coincide with the vacuum of Eq.~\eqref{H4}, as discussed in the introduction. The requirement that physical observables remain finite after renormalization imposes constraints on the UV structure of the state that are incompatible with such a choice.
For instance, in the case of a real scalar field, one-loop consistency requires the state to take the form of a nontrivially squeezed vacuum~\cite{Berezhiani:2023uwt}, and a similar property is expected in the case of a complex scalar.

The classical relation between 
$\mu$ and 
$v$ in \eqref{clsuperfluid} is expected to receive quantum corrections, and therefore does not apply directly in \eqref{intro:coherentsupefluid}. This situation is reminiscent of the spontaneous symmetry breaking in the vacuum, where quantum effects à la Coleman–Weinberg modify the vacuum expectation value. Our goal will be to show that there exists a suitable definition of 
$|\tilde{\Omega}\ra$ such that quantum effects only serve to renormalize this relation, while the time evolution of the state remains consistent with the classical background \eqref{clsuperfluid}. Hence, we keep 
these variables in \eqref{intro:coherentsupefluid} as independent parameters for now, and we fix their relation in the paper by mandating the quantum stationarity of the superfluid configuration.

There are two equivalent possibilities here. One option is to explicitly construct the non-Gaussian state $|\tilde{\Omega}\rangle$.
Alternatively, we could apply the background field method outlined in the introduction and define a superfluid state \eqref{intro:coherentsupefluid} as a sort of a vacuum for fluctuations around the quantum analog of the background \eqref{clsuperfluid}. 
In this case, the decomposition of the form $\hat{\Phi}\equiv ve^{\ri\mu t}+\hat{\psi}$ would lead to the explicitly time-dependent coefficients for the theory of $\hat{\psi}$. However, as we know from semi-classical analysis, these time dependencies are spurious and can be removed by a time-dependent field redefinition. This will also make it easier to define the ground state of the superfluid that is impervious to the flow of time.

To prove that such a state exists, we perform the following time-dependent change of variables generated by the $U(1)$ charge $\hat{Q}$ in the fundamental theory \eqref{H4}
\beq
{\hat{\Phi}=\hat{U}\hat{\varphi}\hat{U}^\dagger = e^{\ri \mu t}\hat \varphi(x)}\,,\qquad \hat{\Pi}=\hat{U}\hat{\Pi}_\varphi\hat{U}^\dagger\,, \qquad {\rm with}\qquad \hat{U}\equiv e^{{\rm i}\mu t\hat{Q}}\,.
\eeq
This parametrization is performed even before introducing the background, and coincides with the field reparametrization that removes the time dependence in the theory of fluctuations.

The new variables $\hat{\varphi}$ and $\hat{\Pi}_\varphi$ are governed by the transformed Hamiltonian \footnote{In general, for a time-dependent transformation implemented on the field content by a unitary operator of the form
\begin{equation}
  \Phi\to \varphi=  e^{\ri \alpha(t) {T}}\Phi e^{-\ri \alpha(t) {T}}\,,
\end{equation}
the Hamiltonian transforms as
\begin{equation}
    H[\Phi] \to \tilde{H}[\varphi]=H[\varphi]-\partial_t({\alpha}(t) T[\varphi])\,.
\label{eq:generalHmodified}
\end{equation}}
\beq
\hat{H}'=\hat{H}-\mu\hat{Q}=\int \rd^3x \left( |\hat{\Pi}_\varphi-{\rm i}\mu\hat{\varphi}^\dagger|^2 +|\vec{\nabla}\hat{\varphi}|^2+\left(m^2-\mu^2\right)|\hat{\varphi}|^2+\lambda |\hat{\varphi}|^4\right)\,.
\label{Hprime}
\eeq
At this point, $\mu$ is still a free parameter. A key feature of this Hamiltonian is that it is time-independent and bounded from below. As such, it has a well-defined vacuum state $|v\ra$ of $\hat{H}'$ which spontaneously breaks $U(1)$ symmetry and carries nonzero charge density. With a usual choice of the phase, the vacuum expectation value can be pointed to the real direction in the field space
\beq
\la v|\hat\varphi|v\ra=\frac{1}{\sqrt{2}}v\,,
\eeq
where $v$ is finally fixed in terms of $\mu$ uniquely. 
The relation between these variables that provides the stationary configuration in the full theory is given in the main body of the paper. 

The fact that \eqref{Hprime} possesses true vacua for every value of $\mu$ is due to the well-known fact that the kinetic matrix of (stable) perturbations around $|v\ra$ can be diagonalized with time-independent wavefunction coefficients, which will be discussed in the 
next section. Furthermore, after a further canonical transformation
\begin{equation}\label{eq:fluctuation_can_trans}
    \hat{\varphi}(t)=v+\hat h+\ri\hat \pi\,.
\end{equation}
we will perform the explicit perturbative construction of $|v\ra$ from the vacuum of the (Lorentz-violating) free theory of superfluid perturbations, using the M${\slashed o}$ller operator. 

The implication of the existence of such a state to all order in perturbation theory is that all equal-time correlation functions $\la v|\hat{\varphi}(x)\hat{\varphi}(y)\ldots|v\ra$ are time-independent.

It is essential to realize that the state $|v\ra$ is only a vacuum for $\hat{H}'$, while having a time-evolution with respect to $\hat{H}$. Otherwise, we would not be able to reproduce the classical evolution given in \eqref{clsuperfluid}. From the point of view of the original operator $\hat{\Phi}$, the state in question is a coherent state, which by definition is not an eigenstate of the Hamiltonian and must evolve in time. In fact
\beq
e^{-{\ri}\hat{H}t}|v\ra=e^{-\ri E_\mu t}e^{-\ri \mu t\hat{Q}}|v\ra\,,
\eeq
where $E_\mu$ is the background energy, which is infinite in the infinite volume limit. In other words, there exists a one-parameter family of coherent states parameterized by $\mu$, corresponding to a different $U(1)$ charge density, the exact time-evolution of which corresponds to the flow in the direction of symmetry. This phenomenon was dubbed as Spontaneous Symmetry Probing in \cite{Nicolis:2011pv} (see also \cite{Nicolis:2012vf, Nicolis:2013sga}).

As a result, the time-evolution of equal-time correlators goes as 
\beq
\la v|\prod_{i=0}^{N_1}\hat{\Phi}(\vec{x}_i,t)\prod_{j=0}^{N_2}\hat{\Phi}^\dagger(\vec{x}_j,t)\ldots|v\ra=e^{\ri(N_1-N_2)\mu t}\la v|\prod_{i=0}^{N_1}\hat{\Phi}(\vec{x}_i,0)\prod_{j=0}^{N_2}\hat{\Phi}^\dagger(\vec{x}_j,0)\ldots|v\ra\,,
\label{evolutionCorrelators}
\eeq
due to the fact that $\hat{Q}$ simply induces the phase-transformation. This shows that the quantum state tracks the classical evolution at all orders. In particular the one-point function evolves as: 
\begin{equation}
    \langle v|\Phi(\vec{x},t)|v\rangle=\frac{v}{\sqrt{2}}e^{\ri \mu t}\,,
\end{equation}
Later in the paper we shall discuss how this behaviour is less trivial than it may seem at first sight, and generically fails for other putative choices of superfluid states with the same one-point function at initial time, once one goes beyond the linearized theory at higher loop order.

It is important to emphasize that both the above discussion and the characterization of the stable state are sensitive to the presence of additional fields charged under the same $U(1)$ symmetry. For instance, if one were to introduce another complex scalar field in such a way that allowed the conversion of $\Phi$-particles into this new species, then a pure superfluid of $\Phi$-particles would be unstable \cite{Cohen:1986ct}.\footnote{For example, one can introduce an additional complex scalar field $\psi$ coupled to $\Phi$ by the interaction term
\begin{equation}
\mathcal{L}\supset g({\hat{\psi}^{*\,2}}\hat{\Phi}^2+ {\hat{\psi}^{2}}\hat{\Phi}^{*\,2})
\end{equation}
to the theory \eqref{eq:LagrSF}.
Although \eqref{clsuperfluid} is still a solution to the classical equation of motion of this modified theory, it would not be quantum stable due to a conversion of $\Phi$ constituents into $\psi$ particles. Stability requires assigning a nonvanishing one-point function to $\hat{\psi}$ as well, to equilibrate the two configurations.}
Instead, if any, the stable state would correspond to an admixture of the two species.

In conclusion, in this work we show that the non-excited superfluid state, which can be thought of as the vacuum for perturbations around the superfluid background and a 
reminiscence of the non-relativistic vacuum of Bogoliubov modes, maintains coherence even at the quantum level. The stabilization is due to the $U(1)$ charge conservation: while the real field condensate is prone to quantum depletion, which is caused by particle number changing processes, charge conservation kills these channels when all the condensate particles carry the same global charge.

A central message is that the physics of the stability can be understood both in the language of fluctuation degrees of freedom and in terms of the constituents of the condensate. In terms of perturbations and $\hat{H}'$, it is clear that we are dealing with the vacuum state of the time-independent Hamiltonian system~\eqref{Hprime} and therefore it should be resilient. In terms of the constituents, the system is described by an appropriately dressed coherent state of charged particles with identical charge. 
Since in this case the state is not an eigenstate of the Hamiltonian~\eqref{H4},
one could envision that particles could re-scatter to form a different superposition from the original coherent state, and in this way leading to quantum breaking. Still, this is not possible since particles in this configuration are not kinematically allowed to produce non-vanishing momentum states. 
However, the stable superfluid state that we have identified is not merely a vanilla coherent state such as~\eqref{intro:coherent}; it comes with an entourage of very specific squeezing and even non-Gaussian adjustments of the vacuum state over which the coherent state is built. And, although a generic coherent state would undergo quantum loss of coherence, the state we studied endures.

\section{A quantum description of superfluids in the background field method}
\label{sec3}
 
In this section, we revisit the standard semiclassical perturbation theory for superfluids. Although the results reviewed here are well known to experienced readers, we show that they nonetheless provide useful insights into the interplay between background stability and the choice of the initial quantum state. We begin by reviewing the classical and semiclassical limits, and then introduce the background field method in the context of superfluids.

\subsection{Classical and semiclassical superfluids}

As we have already discussed in the introduction, classical superfluid configurations are provided by the family of functions
\begin{equation}
\label{classical_superfluid}
\Phi_{\scriptscriptstyle\rm SF}(t)=\frac{v}{\sqrt{2}} e^{{{\rm i} \mu_\text{c} t}}\,,
\end{equation}
with 
\begin{equation}
    \mu_\text{c}^2=m^2+\lambda v^2\,.
    \label{classicalmu}
\end{equation}
Physically, these solutions describe condensates of scalar particles with a net charge density.\footnote{In contrast, zero charge condensates have a null global $U(1)$ charge. In the classical limit, these condensates are described by the following field configuration
\begin{equation}
\Phi_\text{Q=0}(t)=\frac{v}{\sqrt{2}}\text{cd}\left({\sqrt{m^2+\frac{\lambda v^2}{2}}}t,\frac{-1}{1+\frac{4 m^2}{\lambda v^2}}\right)\,,
\end{equation}
where $\text{cd}$ is one of the Jacobi elliptic functions. Here, we set up the field along the real direction. These configurations are not stable at the quantum level, and their long-term dynamics will closely follow that of a real scalar condensate \cite{Berezhiani:2021gph}.}
The parameter $\mu_\text{c}$ plays the role of a classical chemical potential: it sets the energy required to add a particle to the system. Its value is determined by the amplitude $v$, which in turn controls the charge density of the superfluid via the relation $n_{\scriptscriptstyle\rm SF} =  \mu v^2$. As a result, this background spontaneously breaks the global $U(1)$ symmetry of the theory.

To tackle the quantum properties of this background, we start with the semiclassical approach, which consists of perturbing the \textit{classical solution} and quantizing the resulting fluctuations. We perform the following field redefinition
\begin{equation} \Phi(x)=\frac{1}{\sqrt{2}} \Big\{v+h_c(x)+\ri \pi_c(x)\Big\}e^{{\rm i} \mu_\text{c} t}\,,
\label{eq:clPT}
\end{equation}
where $\mu_\text{c}$ is provided by \eqref{classicalmu}. We refer to $h_c(x)$ and $\pi_c(x)$ as the real and imaginary fluctuations. Importantly, this change of variables is exact as it is simply a linear reparameterization of the original field $\Phi(x)$. We use the subscript ‘$c$’ to distinguish these semiclassical fluctuation fields from others that will appear later and be employed throughout the rest of the work.

In the full quantum theory, the fields $h_c(x)$ and $\pi_c(x)$ defined above do not represent proper quantum fluctuations, as their expectation values do not vanish on the vacuum state describing the unperturbed superfluid.  This is because, in an interacting quantum field theory, the expectation value  $\langle\Phi(x)\rangle$ typically does not coincide with the classical solution of the equations of motion, but instead receives quantum corrections. Consequently, consistency requires that $h_c$ and $\pi_c$ acquire nonzero expectation values, for the identity \eqref{eq:clPT} to be satisfied. The presence of such expectation values does not, however, necessarily signal an instability of the background; rather, it may simply indicate that the stable configuration of the full quantum theory differs from its classical counterpart.
We will return to this point in the following sections.

There is a regime in which the classical solution is a proper vacuum for the fluctuation fields, such that their expectation value vanishes. This is the \textit{semiclassical limit}, which is defined by the double scaling limit
\begin{equation}
\lambda \to 0,\qquad\frac{\lambda v^2}{\mu_c^2}\to \text{const}\,.
\label{eq:semiclassicalLimit}
\end{equation}
On the computational side, this semiclassical regime corresponds to truncating correlation functions of the fluctuation fields at tree level, as loop corrections are suppressed in the same limit.
The way this limit emerges will be rigorously examined in the next sections. 

We complete this discussion on the semiclassical theory by studying the theory of the fluctuations $h_s$ and $\pi_s$, in the limit \eqref{eq:semiclassicalLimit}. 
We plug the field redefinition into~\eqref{eq:LagrSF}, and we derive the conjugate momenta
\begin{equation}
\Pi^\pi_{ c} = \dot{\pi}_c + \mu_{\text{c}} h_c, \qquad
\Pi^h_{c} = \dot{h}_c - \mu_{\text{c}} \pi_c\,,
\end{equation}
and the total Hamiltonian density
\begin{equation}
{\mathcal{H}} = \frac{1}{2} \left\{ (\Pi_c^h + \mu_{c} \pi_c)^2 + \partial_i h_c \partial^i h_c + (\Pi_c^\pi - \mu_{\text{cl}} h_c)^2 + \partial_i \pi_c \partial^i \pi_c \right\} + \frac{\lambda}{4} \left( h_c^2 + \pi_c^2 + 2 v h_c \right)^2\,.
\label{eq:Hc}
\end{equation}
We quantize the theory using standard canonical quantization, following the procedure of \cite{Achucarro:2010jv, Creminelli:2023kze,Hui:2023pxc}. First, we impose the standard equal-time canonical commutation relations, and
then we identify the vacuum state of fluctuations, $|\Omega_{h,\pi}^s\rangle$, as the lowest-energy eigenstate of the above Hamiltonian. Note that, although the field redefinition \eqref{eq:clPT} is explicitly time-dependent, the resulting Hamiltonian for the fluctuation fields remains time-independent. This allows for a consistent identification of the vacuum state for fluctuation fields at all times.
With this, we derive the tree-level spectrum of perturbations around the vacuum state $|\Omega_{h,\pi}^s\rangle$, determined according to the relation~\cite{Dodelson}
\begin{equation}
({\omega}^2_{\pm})_c=k^2+\lambda v^2+2 \mu_\text{c}^2\pm \sqrt{4 k^2 \mu_\text{c}^2+\left(\lambda v^2+2\mu_\text{c}^2\right)^2}\,.
\label{eq:semiclassicaldisp}
\end{equation}
As we mentioned, this spectrum is exact in the semiclassical limit~\eqref{eq:semiclassicalLimit}, and it is corrected only once the above limit is relaxed.
Moreover, since the configuration~\eqref{classical_superfluid} spontaneously breaks the underlying $U(1)$ invariance, the theory admits a gapless fluctuation mode \cite{Goldstone:1962es,Nielsen:1975hm} given by the negative branch of~\eqref{eq:semiclassicaldisp}. Notice that both branches of the spectrum are stable, which is important for the identification of the vacuum. This is in contrast to the case of attractive bosons, where the negative branch corresponds to an unstable sound mode \cite{Guth:2014hsa}.

In the low-momentum limit, the dispersion relation of the gapless mode reduces to that of a sound wave
\begin{equation}
    \omega_-\simeq c_s k\qquad \text{with}\qquad c^2_s=\frac{\lambda v^2}{2\mu_c^2}\,.
    \label{eq:phononspectrum}
\end{equation}
This corresponds to the phonon spectrum, which represents the sole degree of freedom governing the low-energy dynamics of the superfluid.\footnote{The massive mode does not participate in the dynamics at energies well below its gap, approximately set by $\mu_c$.} In the non-relativistic limit, obtained by setting $\mu_c\sim m$, the parameter $c_s$ coincides with the sound speed of the fluid. By comparing with~\eqref{eq:semiclassicalLimit}, we see that the semiclassical limit corresponds to sending the coupling to zero, while keeping the sound speed finite.

\subsection{Quantum superfluids and the background field method}
\label{sec:Timeind}
\begin{figure}
\includegraphics[width=1\textwidth, trim=20 650 20 10, clip]{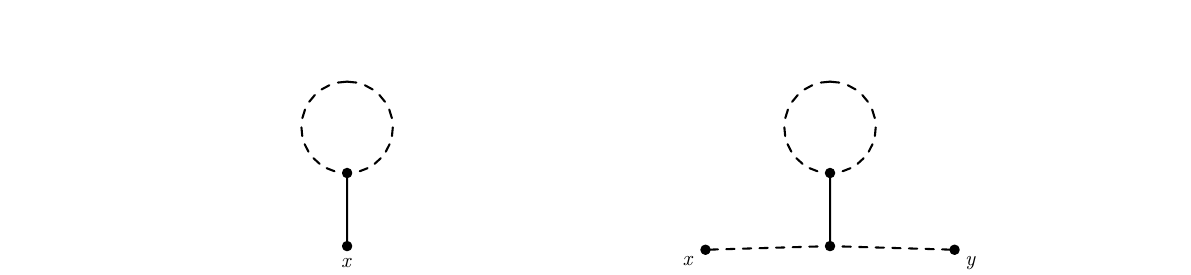}
\caption{Example of tadpole diagrams. The first corresponds to a correction to the one-point function of the fluctuation field $h$. The second is a reducible diagram that appears in the computation of loop corrections to the $\langle \pi(x) \pi(y)\rangle$ correlation function. Diagrams of this type signal an incorrect choice of background and can be consistently reabsorbed into a shift from the classical solution to the quantum-corrected one. }
\label{fig:fig1}
\end{figure}
If the semiclassical limit~\eqref{eq:semiclassicalLimit} is relaxed, the condition $\langle h_c(x) \rangle = \langle \pi_c(x) \rangle = 0$ cannot be consistently maintained beyond the tree level. This can be immediately seen by expanding the full theory \eqref{eq:Hc} perturbatively around the quadratic theory. 
For example, consider the cubic vertex $h \pi^2$ in the Hamiltonian~\eqref{eq:Hc}. This induces the following one-loop correction to the expectation value of the real fluctuation
\begin{equation}
    \langle h(x)\rangle\supset-\lambda^2 v^2\,\text{Im}\int_{-\infty_-}^t \rd^4z \langle \pi_{c,0}^2(z)\rangle\langle h_{c,0}(z)h_{c,0}(x)\rangle\,,
\end{equation}
where $h_{c,0}(x)$ and $\pi_{c,0}(x)$ are the fluctuation fields of the quadratic theory.
Here, we applied the interaction picture formalism~\cite{Peskin:1995ev,Weinberg:2005vy, Adshead:2008gk}, which we illustrate in Section~\ref{section:interaction_picture}.
The above correction is pictured as a tadpole diagram, represented by the left panel of Fig.~\ref{fig:fig1}, and signals that the expansion is organized around the incorrect background. Moreover, tadpoles propagate in loop corrections to all correlation functions, in the form of one-particle-reducible diagrams as shown in the right panel of Fig.~\ref{fig:fig1}. This class of diagrams cannot be neglected, as tadpoles contain physical information about the quantum background and conspire to keep the gapless mode massless beyond the tree-level spectrum.

However, the presence of tadpoles is highly inconvenient from a computational standpoint. At each order in perturbation theory, they provide a set of diagrams that must be resummed through a shift of the background field itself. In this section, we present an improved organization of the perturbative expansion in which such contributions are automatically accounted for. This is the improved version of the \textit{background field method}~\cite{Baacke:1997zz, Berezhiani:2021gph}.

We begin by introducing the state $|v\rangle$, the \textit{unperturbed superfluid state}, whose properties we systematically develop throughout this and the following section. A summary of its defining features, along with an explicit representation, is provided in Section~\ref{sec:state}. 
We perform the following parameterization of the scalar field
\begin{equation}
\Phi(x)=\frac{1}{\sqrt{2}}\left(v + h(x) + \ri \pi(x)\right)e^{\ri \mu t}\,.
\label{eq:SFdecomposition}
\end{equation}
Here, $h(x)$ and $\pi(x)$ denote the \textit{real} and \textit{imaginary} fluctuation fields, while $v$ and $\mu$ are parameters. By considering the expectation value of the above relation over the state $|v\rangle$, we then impose the tadpole conditions
\begin{equation}
\langle h(x)\rangle = \langle \pi(x)\rangle = 0 \qquad\Longrightarrow \qquad \langle \Phi(x)\rangle = \frac{v}{\sqrt{2}} e^{\ri \mu t}\,,
\label{eq:tadpoleC}
\end{equation}
which ensure that $|v\rangle$ serves as a vacuum for the fluctuations~\cite{Boyanovsky:1998aa,Berezhiani:2021gph} also beyond the limit~\eqref{eq:semiclassicalLimit}.

Although this decomposition resembles the semiclassical expansion introduced in the previous section, it is conceptually different: here, the relation between $v$ and $\mu$ will be determined by the requirement that the background remains stable under quantum corrections, and not by the classical equation of motion. 
The only assumption we make at this stage is that both $v$ and $\mu$ are time-independent, as we are interested in a stable quantum superfluid configuration. In particular, under this assumption, the one-point function of the field operator, Eq.~\eqref{eq:tadpoleC}, exhibits the same time dependence as the classical solution~\eqref{classical_superfluid}, with the classical parameter $\mu_c$ replaced by the quantum analog $\mu$.
We will show that this assumption is consistent, provided the state  $|v\rangle$  is appropriately defined. In other words, there exists a state for which a constant amplitude and chemical potential solve the full quantum Heisenberg equations of motion.\footnote{The decomposition~\eqref{eq:SFdecomposition} thus corresponds to a restricted subclass of possible parameterizations. The most general form is given by
\begin{equation}
\Phi(x) = \frac{1}{\sqrt{2}} \left\{ \Phi(t) + h(x) + \ri \pi(x) \right\} e^{\ri \omega(t)}\,,
\end{equation}
with initial conditions $\Phi(0) = v$, $\omega(0) = 0$, $\dot{\omega}(0) = \mu$, and $\dot{v}(0) = 0$.  We aim to show that there exists a definition of $|v\rangle$ for which $\omega(t)=\mu t$ and $v=\text{const}$ is a consistent solution of the Heisenberg field equation. }
Finally, the condition~\eqref{eq:tadpoleC} fixes the initial condition for the one-point function of the field operator and its conjugate momentum at the time of definition of the state, with the first set by $v$ and the second by $\mu$.

We move to determining the relation between the amplitude and the chemical potential in the full quantum theory.
The dynamics of the theory is encoded in the equation of motion of the Heisenberg field operator $\Phi(x)$:
\begin{flalign}
\square \Phi+m^2\Phi+2{\lambda}|\Phi^2|\Phi=0\,.
\label{eq:EOMSF}
\end{flalign}
We substitute the field redefinition \eqref{eq:SFdecomposition} in the above equation, still assuming that $v$ and $\mu$ are time-independent functions. 
We obtain a set of two equations by isolating the real and imaginary parts of the resulting equation:
\begin{gather}
\left(\square -\mu^2+m^2+3\lambda v^2 \right)h-2\mu\dot{\pi}+\left(m^2+\lambda v^2-\mu^2\right)v +\lambda v (3h^2+\pi^2)+\lambda\left(h^3+h \pi^2\right)=0 \,,
\label{eq:radialEqCondensate}\\
\left(\square-\mu^2+m^2+\lambda v^2\right)\pi+2\mu\dot{h} +\lambda\left(
\pi^3+ h^2 \pi+2 vh \pi \right)=0 \,.
\label{eq:angularEqCondensate}
\end{gather}
In this approach, the Heisenberg equations of motion can be split into a set of coupled equations for the one-point function of the field operators in the superfluid state and for the higher order $n$-point functions for the fluctuation fields \cite{Berezhiani:2021gph}. 

The first set of equations is obtained by considering the expectation value of Eq.~\eqref{eq:radialEqCondensate} and Eq.~\eqref{eq:angularEqCondensate} over $|v\rangle$, and using the tadpole conditions:
\begin{flalign}
&\mu^2=m^2 +\lambda v^2+3\lambda \langle h^2\rangle+\lambda   \langle \pi^2\rangle+\frac{\lambda}{v}\langle h^3\rangle+ \frac{\lambda}{v}\langle h \pi^2\rangle\,,
\label{eq:radialU1}\\ &\langle h\pi\rangle+\frac{1}{2v}\langle \pi h^2+\pi^3\rangle =0\,.
\label{eq:angularU1}
\end{flalign}
Eq.~\eqref{eq:radialU1} gives us the relation between $v$ and $\mu$ for a stationary condensate at the quantum level. It is the non-perturbative definition of the quantum chemical potential and represents one of the main results of this section. In particular, the one-point function defined by \eqref{eq:tadpoleC}, evaluated along the chemical potential given in \eqref{eq:radialU1}, defines the quantum background which resums all tadpole diagrams that arise when expanding perturbatively around the classical solution as in the previous section. 
In contrast, Eq.~\eqref{eq:angularU1} 
is an identity that correlation functions have to satisfy, and trivializes in the classical limit.

In the next sections, we check that if we solve for the correlation functions of fluctuation fields, the above relations can be satisfied by a constant $\mu$ and constant $v$ provided that the state $|v\rangle$ is chosen appropriately, which is not yet guaranteed at this stage. In particular, we have to understand how fluctuations act on the state $|v\rangle$, and show that there is a choice that makes all the correlation functions entering \eqref{eq:radialU1} time-independent. Notice that \textit{not all choices} of the initial conditions for fluctuations will result in this, as we show in Section~\ref{eq:UnstableSF}. 

The dynamics of the fluctuation fields is governed by a second set of equations, which in turn determines also the behavior of the correlation fluctuations in the fluctuation fields. These are obtained by plugging the relation~\eqref{eq:radialU1} back into~\eqref{eq:radialEqCondensate} and \eqref{eq:angularEqCondensate}, and read
\begin{flalign}
&(\square+M_h^2) h+2\mu\dot{\pi} +\lambda v (3h^2-3\langle h^2\rangle+\pi^2-\langle \pi^2\rangle)+\lambda\left(h^3-\langle h^3\rangle +h \pi^2-\langle h \pi^2\rangle\right)=0\,,
\label{eq:radialEqCondensate2}\\
&(\square+M_\pi^2) \pi+2\mu\dot{h} +\lambda\left(
\pi^3+ h^2 \pi+2 vh \pi\right)=0\,.
\label{eq:angularfl}
\end{flalign}
Here, we defined 
\begin{flalign}
    &M_h^2=2\lambda v^2-3 \lambda\langle h^2\rangle-\lambda\langle\pi^2\rangle-\frac{\lambda}{v}\langle h\pi^2\rangle-\frac{\lambda}{v}\langle \pi^3\rangle\, , \\
    &M_\pi^2= -\lambda\langle \pi^2\rangle-3\lambda\langle h^2\rangle -\frac{\lambda}{v}\langle \pi^3\rangle-\frac{\lambda}{v}\langle h\pi^2\rangle\,.
\end{flalign}

In contrast to the standard vacuum analysis, the operatorial equations of motion here depend explicitly on the expectation value of the operators themselves in the quantum state of the system. These expectation values are evaluated at time and space coincidence, and we show that there exists a choice of the state for which they remain time-independent.
The presence of these additional terms has important consequences: first, taking the expectation value over the state  $|v\rangle$, we find that the one-point functions of the fluctuation fields vanish at all times. Therefore, $\langle h(x)\rangle=\langle \pi(x)\rangle=0$ is a solution\footnote{For the imaginary equation, one must also rely on Eq.~\eqref{eq:angularU1}.} as expected, making the formalism we defined self-consistent. Second, the imaginary mode seems to acquire a 
gap at first glance, which might suggest that the would-be Goldstone boson becomes massive. However, in Section~\ref{sec:gaprel}, we show that these apparent mass terms are canceled by non-linear interactions, ultimately restoring gaplessness. 

Finally, this set of equations is coupled to the one governing the one-point function~\eqref{eq:radialU1}. This coupling arises because the fluctuation fields are defined in terms of the parameter 
$\mu$, which in turn depends on the fluctuation fields themselves. In general, this is not an issue as it is always possible (although technically difficult) to solve this system perturbatively. We refer the reader to the case of a real scalar field \cite{Baacke:1997zz, Berezhiani:2021gph}, where both the equation for fluctuations and for the background are dynamical equations.  However, in the case of a stable superfluid configuration, the situation is considerably simpler: Eq.~\eqref{eq:radialU1} acts as a constraint rather than a dynamical equation and $\mu$ can always be eliminated accordingly, as we demonstrate in the next section.
Therefore, the system of Eqs. \eqref{eq:radialU1}–\eqref{eq:angularfl} fully encodes the quantum dynamics of the background and its fluctuations, under the assumption of a constant amplitude and chemical potential.

How is the semiclassical limit recovered? As previously discussed, it corresponds — computationally — to neglecting loop corrections to the correlation functions. From the perspective of the one-point function, this translates into discarding all contributions from fluctuation fields in Eq. \eqref{eq:radialU1}, implying that the chemical potential $\mu$ reduces to its classical counterpart $\mu_c$. The same logic applies to the equations for the fluctuations: since all expectation values appearing therein are evaluated at the coincidence, they correspond to loop corrections and must likewise be neglected for consistency. As a result, the equations of motion for the fluctuations reduce precisely to those obtained in the semiclassical expansion discussed in the previous section.
\subsection{Solving the equations of motion of fluctuation fields using the interaction picture}
\label{section:interaction_picture}
A key advantage of stable superfluid systems is that the full quantum equation of motion for the background, Eq.~\eqref{eq:radialU1}, acts as a constraint rather than a dynamical equation. This allows us to express the chemical potential $\mu$ in terms of its classical counterpart $\mu_c$ and correlation functions involving the fluctuation fields. We can then substitute this expression back into the equation for fluctuations, effectively eliminating $\mu$.

After plugging $\mu$ in, there are two distinct approaches to solving the equations for the fluctuation fields. The first involves reformulating them as a system of equations for correlation functions. This is done by multiplying Eqs.~\eqref{eq:radialEqCondensate2} and \eqref{eq:angularfl} by arbitrary products of $\pi(y_i)$ and $h(y_j)$, and then taking the expectation value in the state $|v\rangle$. The result is an infinite, coupled hierarchy of ordinary differential equations for correlation functions, commonly referred to as the Schwinger–Dyson equations. An explicit example is provided in Section~\ref{sec:gaprel}, where this approach proves useful in demonstrating the one-loop gaplessness of the phonon.

\begin{figure}
\includegraphics[width=1\textwidth, trim=20 650 20 10, clip]{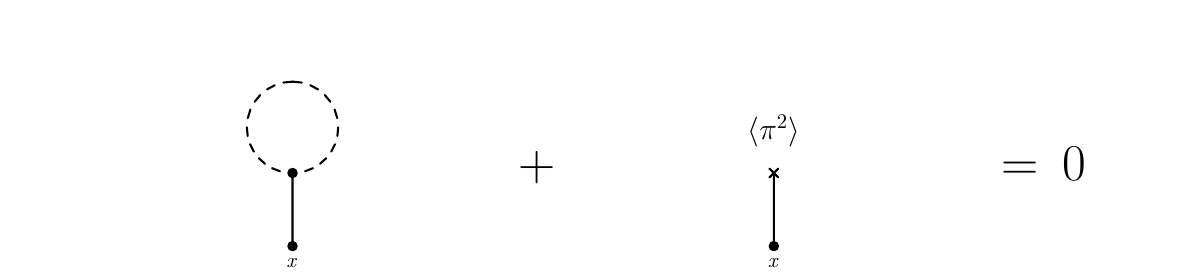}
\caption{In the background field method, tadpoles in the interacting Hamiltonian are cancelled by those arising from nonlinear interaction terms. For instance, in the figure above, the vertex $h\langle \pi^2\rangle$ cancels the tadpole diagram that is generated by the cubic vertex $h\pi^2$.  }
\label{fig:fig2}
\end{figure}

The second approach, which we adopt here, consists of finding an operatorial solution using the interaction picture formalism. This method is fully equivalent to the Schwinger–Dyson approach, as the resulting operatorial solution can be used to construct all correlation functions that would otherwise be derived from the Schwinger–Dyson hierarchy. 
To do so, let us derive the Hamiltonian for fluctuation fields. We begin by inserting the field redefinition~\eqref{eq:SFdecomposition} into the fundamental Lagrangian~\eqref{eq:LagrSF}, from which we calculate the canonical momenta
\begin{equation}
\Pi^h = \dot{h} - \mu \pi\,, \qquad \Pi^\pi = \dot{\pi} + \mu h\,.
\end{equation}
We then impose the canonical equal-time commutation relations,
\begin{equation}
[h(\vec{x},0), \Pi^h(\vec{y},0)] = [\pi(\vec{x},0), \Pi^\pi(\vec{y},0)] = \ri \delta(\vec{x}-\vec{y})\,,
\end{equation}
with all other commutators vanishing. Finally, we derive the full Hamiltonian density  $\mathcal{H}=\mathcal{H}_0+\mathcal{H}_\text{int}$,
where
\begin{flalign}
&\mathcal{H}_0=\frac{1}{2}\left\{(\Pi^\pi-\mu h)^2+\partial_i\pi \partial^i \pi+(\Pi^h+\mu \pi)^2+\partial_i h \partial^i h+2\lambda v^2h^2\right\}\,,
\label{eq:H2}
\\&\mathcal{H}_\text{int}=\frac{\lambda}{4}\left(\pi^2+h^2+4 v h\right)(\pi^2+h^2)-\frac{\lambda}{2}(\pi^2+h^2+2v h)\left(3\langle h^2\rangle+\langle \pi^2\rangle+\frac{1}{v}\langle h^3\rangle+\frac{1}{v}\langle h\pi^2\rangle\right)\,.
\label{eq:Hint}
\end{flalign}
A few remarks are in order regarding the structure of the Hamiltonian.
First, terms proportional to $\mu^2$ in the Hamiltonian~\eqref{eq:Hint} have been eliminated using the background equation~\eqref{eq:radialU1}.  Also, we observe the presence of tadpole terms in $\mathcal{H}_\text{int}$. These are precisely cancelled by the contributions from tadpole diagrams generated by cubic and higher-order interactions, see Fig.~\ref{fig:fig2}. A similar cancellation occurs for the apparent mass term of the Goldstone mode: while $\mathcal{H}_\text{int}$ includes terms quadratic in $\pi$, which suggest a spurious gap, we will show that these are cancelled once loop corrections from higher-order vertices are consistently included.\footnote{The decomposition of the full Hamiltonian into free and interacting parts is, in principle, arbitrary. For instance, one could rewrite the chemical potential $\mu$ appearing in the free Hamiltonian~\eqref{eq:H2} in terms of 
 $\mu_c$ and expectation values of the fluctuation fields, and move the latter into $\mathcal{H}_\text{int}$. However, such a splitting would be inconvenient, as $\mu$ (and not $\mu_c)$ is the physical, finite (after renormalization) parameter that characterizes the superfluid in the full quantum theory. A similar remark applies to the linear and quadratic terms in  $\mathcal{H}_\text{int}$: while they could in principle be absorbed into $\mathcal{H}_0$, they are designed to cancel tadpole contributions when combined with the non-linear vertices, and are thus best kept within the same interaction sector. }  

The quadratic part $\mathcal{H}_0$ matches the structure of the semiclassical Hamiltonian~\eqref{eq:Hc}, with the only difference being the replacement of the classical chemical potential $\mu_c$ with its fully quantum counterpart $\mu$. This identification allows us to introduce the interaction picture as in Refs.~\cite{Achucarro:2010jv,Creminelli:2023kze,Hui:2023pxc}. This will be useful to analyse the dynamics of the superfluid state.

We introduce the interaction picture fields, as the fields evolving only under the time evolution set by $H_0$. These are related to the Heisenberg picture fields by 
\begin{equation}
    \pi(x)= U^\dagger(t,t_*)\pi_0(x)U(t,t_*)\,, \qquad h(x)= U^\dagger(t,t_*) h_0(x)U(t,t_*)\,,
\label{eq:interaction}
\end{equation}
Here, $t_*$ is the standard fiducial time at which the interaction and Heisenberg picture coincide. The operator $U(t_1,t_2)$ is the interaction picture evolution operator, defined as
\begin{equation}
    U(t_1,t_2) =\mathcal{T}\,\text{exp}\left\{{-{\rm i}\int_{t_2}^{t_1} } {\rm d}^4z \,\mathcal{H}_\text{I}(z)\right\}\,,
\end{equation}
where $\mathcal{H}_\text{I}(z)=\mathcal{H}_\text{int}[h_0(z),\pi_0(z)]$ and with $d^4z=dz^0 d^3z$.
To carry out explicit computations, we have to decompose $h_0(x,t)$ and $\pi_0(x,t)$ in terms of a ladder expansion. This defines the vacuum $|0_{h, \pi}\rangle$ of the quadratic Hamiltonian $\mathcal{H}_0$.

According to the quadratic Hamiltonian \eqref{eq:H2}, the real and imaginary fields are kinetically mixed. Therefore, we cannot introduce two independent sets of ladder operators for $h_0(x)$ and $\pi_0(x)$. However, if the quadratic theory is diagonalized, two independent degrees of freedom emerge, with a dispersion relation determined by \eqref{eq:semiclassicaldisp}, albeit with $\mu_\text{cl}$ replaced by $\mu$.
Following \cite{Achucarro:2010jv,Creminelli:2023kze, Hui:2023pxc}, we introduce the ladder operators for the diagonal modes, defined as $a_\alpha(k)$ with $\alpha=\pm$, and impose that they satisfy  the following commutation relations 
\begin{equation}
    [a_\alpha (\vec{k}), a^\dagger_\beta(\vec{k}')]= (2\pi)^3 \delta_{\alpha \beta} \delta^{(3)}(\vec{k}-\vec{k}')\,.
    \label{aadagger}
\end{equation}
Then, the vacuum $|0_{h,\pi}\rangle$ is defined in such a way that it is annihilated by $a_\alpha(\vec{k})$ for every vector $\vec{k}$.
Having defined the ladder expansions of the diagonal modes, we use them as a basis to expand the free real and imaginary quantum fields:
\begin{flalign}
&\phi^a_0(x)=\sum_{\alpha=+,-}\int \frac{\rd^3k }{(2\pi)^3 \sqrt{2 \omega_\alpha}} \left(a_\alpha Z_{\alpha}^a e^{-\ri\omega_\alpha t+\ri \vec{k} \cdot \vec{x}}+\text{h.c.}\right)
\label{eq:condensateDecomposition}\,, \\ 
&\Pi_0^a(x)= \sum_{\alpha=+,-}\int \frac{\rd^3k }{(2\pi)^3 \sqrt{2\omega_\alpha}} \left(\left(-\ri \omega_\alpha Z_{\alpha}^a+\mu\epsilon_{\alpha \beta}Z_{\beta}^a \right)a_\alpha  e^{-\ri\omega_\alpha t+\ri \vec{k} \cdot \vec{x}}+\text{h.c.} \right)\,,
\label{eq:condensateDecomposition2}
\end{flalign}
with $\phi^1=\pi$ and $\phi^2=h$. The wavefunction coefficients required for the compatibility of \eqref{aadagger} with canonical commutation relations read
\begin{flalign}
    Z_\pm^\pi(k)=(-{\rm i})^{\frac{1}{2}\pm \frac{1}{2}}\sqrt{\frac{1}{2}\pm\frac{2 \mu^2-\lambda v^2}{2\sqrt{4 k^2 \mu^2+(\lambda v^2+2\mu^2)^2}}} \,,
      \label{eq:Zpi}\\
    Z_\pm^h(k)=(-{\rm i})^{\frac{1}{2}\mp \frac{1}{2}}\sqrt{\frac{1}{2}\pm\frac{2 \mu^2+\lambda v^2}{2\sqrt{4 k^2 \mu^2+(\lambda v^2+2\mu^2)^2}}} \,.
    \label{eq:Zh}
\end{flalign}

Observe that in contrast to the vacuum theory, the wavefunction coefficients $Z^a_\alpha$ are non-trivial in the case of the superfluid background, due to the kinetic mixing. 
We refer the reader to the aforementioned references for the derivation of the above ladder expansion. 

Using the interaction picture formalism, (equal-time) correlation functions of Heisenberg picture fluctuation fields can be expressed in terms of interaction-picture fields as
\begin{equation}
     \langle v| \mathcal{O}[h,\pi] |v \rangle= \langle v| U^\dagger(t,t_*) \mathcal{O}[h_0,\pi_0]  ~ U(t,t_*)|v\rangle\,,
     \label{eq:ExpFl}
\end{equation}
the evaluation of which, in turn, requires an explicit form for the state.

\subsection{Stable superfluids and initial conditions}
\label{sec:state}

We conclude this section by presenting the explicit expression for the stable superfluid state. Up to this point, we have derived the equations of motion under the assumption that a hypothetical state $|v\rangle$ exists, characterized by a constant amplitude and a time-independent chemical potential.

In what follows, we show that this state is uniquely identified with the interacting vacuum of the fluctuation Hamiltonian (the sum of Eqs.~\eqref{eq:H2} and \eqref{eq:Hint}).
The explicit expression of the interacting vacuum is:
\begin{equation}
|v\rangle \equiv |\Omega_{h,\pi}\rangle = U(t_*, -\infty_-) |0_{h,\pi}\rangle\,,
\label{eq:non-gaussianV}
\end{equation}
with $\infty_\pm=\infty(1\pm \ri \epsilon)$.
The second equality, valid up to a normalization factor, expresses the interacting vacuum of fluctuations in terms of the free vacuum $|0_{h,\pi}\rangle$ — the ground state of the quadratic Hamiltonian~\eqref{eq:H2}.\footnote{
It represents the standard projection of the interacting vacuum 
on the free vacuum (see chapter~4.2 of Ref.~\cite{Peskin:1995ev}). }
Such a state corresponds to the superfluid defined in Section \ref{sec:state_def}, as it is evident by performing the time-independent canonical transformation \eqref{eq:fluctuation_can_trans}.

Next, we prove that the state in question has the following properties
\begin{enumerate}[label=(\roman*)]
    \item Equal-time correlation functions of the fluctuation fields $h(x)$ and $\pi(x)$ in their interacting vacuum are time-independent.
    \item Such correlation functions 
    containing an odd power of $\pi(x)$ are vanishing.
\end{enumerate}
The property (i) is enough to guarantee that the chemical potential $\mu$ is constant according to Eq.~\eqref{eq:radialU1}, while property (ii) implies that Eq. \eqref{eq:angularU1} is trivially satisfied. Hence, the two properties are enough to guarantee that the background described by this state is stationary.

Let us begin with property (i), which follows directly from the time-independence of the full Hamiltonian density governing the fluctuation theory. In particular, by specifying the explicit form of the state $|v\rangle$, we can now set initial conditions and rewrite any expectation value involving the fluctuation fields, as in equation~\eqref{eq:ExpFl}, in the form
\begin{flalign}
  \langle v| \mathcal{O}[h,\pi] |v \rangle=&  \langle 0_{h,\pi}| U^\dagger(t,-\infty_-) \mathcal{O}[h_0,\pi_0] ~ U(t,-\infty_-)|0_{h,\pi}\rangle\,,
\label{eq:correlatorStable}
\end{flalign}
The expression above reduces to a standard in-in correlator evaluated on a stable vacuum. It leads to time-independent correlation functions at point-coincidence, provided two conditions are met: the mode functions of the free theory have time-independent wavefunction coefficients, and the coupling constants entering the interaction Hamiltonian are time-independent.
The first condition is straightforwardly verified by inspecting the explicit form of the free-theory solutions~\eqref{eq:condensateDecomposition}. The second, however, requires more care. In the background field method, the interaction Hamiltonian contains coefficients that depend on equal-time correlation functions. As such, assuming their time-independence from the outset would be tautological.
Nonetheless, in a perturbative loop expansion, the coefficients appearing in the interaction terms at a given loop order are determined solely by correlators from lower orders. By recursion, as long as all tree-level correlators are time-independent, no spurious time dependence arises at any finite order in perturbation theory.

Let us now address property (ii). In the case of spontaneous symmetry breaking in a vacuum, correlators involving an odd number of imaginary fluctuations vanish due to the underlying $\mathbb{Z}_2$ symmetry of the $\pi$-sector. For a superfluid, however, the situation is qualitatively different: the kinetic mixing between real and imaginary field explicitly breaks this symmetry. This is evident, for instance, from the fact that the mixed correlator $\langle h(x)\pi(y)\rangle$ is nonzero if $t_x\neq t_y$.

Nonetheless, the correlator vanishes when evaluated at equal times, i.e., for $t_x = t_y$. This follows from the fact that the kinetic mixing term is invariant under a combined $\mathbb{Z}_2$ transformation and time reversal:
\begin{equation}
\pi \to -\pi, \qquad t \to -t\,.
\label{eq:symmetryKM}
\end{equation}
As a result, equal-time correlators obey
\begin{equation}
\langle v| h^m(x,t), \pi^n(\vec{x},t) |v \rangle = (-1)^n \langle v| h^m(x,-t), \pi^n(\vec{x},-t) |v \rangle = (-1)^n \langle v| h^m(x,t), \pi^n(\vec{x},t) |v \rangle\,,
\label{eq:timeinvZ2}
\end{equation}
where in the last step we used the time-independence of equal-time correlators, as established in property (i).
If $n$ is odd, Eq.~\eqref{eq:timeinvZ2} implies that correlation functions involving an odd number of $\pi$ vanish when evaluated at the same time.

With this, we thus conclude that a stable superfluid is described by the state minimizing the energy of fluctuations around the one-point function~\eqref{eq:tadpoleC}, which is in turn defined by a constant amplitude $v$ and a phase $\mu$ satisfying Eq.~\eqref{eq:radialU1}.

Finally, let us observe that, even if we have constructed the state using the fluctuation fields $h$ and $\pi$, one can equivalently characterize it in terms of the fundamental field of the theory by undoing the canonical transformation \eqref{eq:fluctuation_can_trans}. This allows one to obtain all the correlation functions of $\Phi$ and its canonical momentum, which completely specify the state. Moreover, this clarifies that the stable superfluid state can be seen in terms of $\Phi$ as a coherent state with a very specific dressing. The dressing is needed to minimize the energy of fluctuations around the one-point function once higher loops are taken into account. For further details, see Appendix~\ref{appstate}.

It is important to emphasize that the stationary superfluid state~\eqref{eq:non-gaussianV} is therefore intrinsically non-Gaussian. As shown in Ref.~\cite{Berezhiani:2021zst, Berezhiani:2023uwt}, the non-Gaussian dressing is in general essential for describing a configuration with a finite energy density. In Section~\ref{eq:UnstableSF}, we will return to this point and show that representing the superfluid state solely as a squeezed coherent state also results in a quantum mechanically unstable configuration. Thus, beyond ensuring the finiteness of observables, the non-Gaussian dressing plays an essential role in guaranteeing stability.
For more details on the form of the correlation functions and on the explicit coherent state construction, we refer to the Appendix \ref{App:squeezing}.

\subsection{The superfluid state as the ground state of $\hat{H}-\mu\hat{Q}$}
\label{sec:SSP}
It is important to remark on the peculiar property of the state $|v\rangle$, that we identified as the stable superfluid background. As explained in Section \ref{sec:state_def}, although it is not an eigenstate of the Hamiltonian due to its explicit time dependence, its evolution is nevertheless remarkably simple: the time dependence of all correlation functions of the field operator $\Phi(x)$ (and conjugate momentum) simply reduces to an overall oscillating phase, identical to that of the one-point function of the field operator.
In particular, this shows that the system retains its classical behaviour indefinitely, without undergoing any quantum depletion.

The simplicity of its time evolution is not accidental, but rather a consequence of the fact that the state is a Spontaneous Symmetry Probing (SSP) state, in the sense of Ref.~\cite{Nicolis:2011pv}. 
Consider the class of states that spontaneously break the internal symmetry of a theory, associated with a conserved charge ${Q}$.  SSP states $|c\rangle$ are defined as those whose time evolution follows the symmetry direction. Formally, these are defined by the relation
\begin{equation}
\left({H}-\mu  Q\right)|c\rangle=0\,.
\label{eq:ModiiedH}
\end{equation}
Thus, while they are not eigenstates of the Hamiltonian, they are eigenstates of the modified operator ${H} - \mu {Q}$. For states satisfying this identity, the dynamics is equivalently controlled by the generator of the symmetry ${Q}$. 
 In particular, the charge operator can only rotate symmetry-breaking 
 states into each other.
 Therefore, because its action over $|c\rangle$ leads to physically equivalent states, so does the Hamiltonian according to \eqref{eq:ModiiedH}. In this sense, coherence is preserved by the time evolution for SSP states.

Since the superfluid state $|v\rangle$ is defined as the vacuum of the time-independent Hamiltonian \eqref{Hprime}, it precisely satisfies the definition of an SSP state.

SSP states for superfluid-like states were constructed in \cite{Nicolis:2011pv} for a linearized theory. In that setting, the authors showed that they take the form of a coherent state. However, a simple coherent state fails to be an SSP state once one goes beyond the linearized theory. In particular, we shall show in the next section how a coherent state with Gaussian squeezing undergoes a non-trivial quantum dynamics at two-loop order, which progressively spoils its classicality by sourcing higher order n-point functions at the expense of the energy originally stored in the one-point function. Such a process, named quantum breaking in \cite{Dvali:2017eba}, spoils the stationarity of the state and consequently its SSP properties.

On the other hand, we showed that the superfluid state $|v\rangle$ defined above is ``stationary'' at all times, and is an SSP state at all orders in the loop expansion. In the next section, we clarify how the specific dressing of the coherent state, or equivalently the specific choice of all the higher-order correlation functions, is paramount to ensure that stability is preserved once quantum corrections are taken into account.
\section{One-loop chemical potential and two-loop (in-)stability}
\label{sec:stability}

In the previous section, we derived the equations governing the stationarity of the superfluid state. Using symmetry-based arguments, we showed that these equations are satisfied by a stationary configuration if the quantum state of the system is identified with the interacting vacuum of the superfluid fluctuations. In this section, we explicitly check this result by evaluating these equations within a perturbative loop expansion.

We start by computing the renormalized one-loop chemical potential. We demonstrate that the divergences that appear can be renormalized away via the standard vacuum renormalization techniques. We then proceed to the two-loop level, showing that both stationarity conditions~\eqref{eq:radialU1} and~\eqref{eq:angularU1} are satisfied by a constant chemical potential $\mu$. Moreover, we show that equal-time correlation functions of fluctuation fields are time-independent. 

Finally, we show that not all states that reproduce the classical behavior of Eq.~\eqref{classical_superfluid} provide a stationary quantum dynamics for the class of systems under consideration. In particular, the highly non-Gaussian structure of the interacting vacuum in Eq.~\eqref{eq:non-gaussianV} plays a crucial role in ensuring stationarity. In the final part of this section, we examine what happens when the interacting (non-Gaussian) vacuum is replaced with a Gaussian state, such as a squeezed coherent state. We show that, starting at two-loop order, the dynamics becomes unstable: the one-point function of the field operator begins to deviate from the stationary profile~\eqref{eq:tadpoleC}. This proves that undressed coherent states fail in describing configurations that track the classical dynamics in the most accurate way possible, in interacting field theories. Moreover, this computation shows how the formalism developed in the previous section — combining the canonical approach with the background field method — can be readily extended to non-equilibrium configurations.

\subsection{The renormalized one-loop chemical potential}

To compute the chemical potential, we evaluate the correlation functions appearing in Eq.~\eqref{eq:radialU1} using the identity given in Eq.~\eqref{eq:correlatorStable}. This relation can be systematically expanded in powers of the interaction Hamiltonian, following the approach of Ref.~\cite{Adshead:2008gk}. One obtains:
\begin{flalign}
\langle v|\mathcal{O}(x)|v\rangle &= \langle \mathcal{O}_0(x)\rangle
- 2\,\text{Im} \int_{-\infty_-}^t \!\!\!\! \rd^4 z_1\, \langle \mathcal{H}_I(z_1)\mathcal{O}_0(x)\rangle \nonumber  + \int_{-\infty_+}^t \!\! \!\! \rd^4 z_1 \int_{-\infty_-}^{t} \!\! \!\!\!\!\rd^4 z_2\, \langle \mathcal{H}_I(z_1)\mathcal{O}_0(x)\mathcal{H}_I(z_2)\rangle \nonumber \\
&\quad - 2\,\text{Re} \int_{-\infty_+}^t \!\!\!\! \rd^4 z_2 \int_{-\infty_+}^{t_1} \!\! \!\!\rd^4 z_1\, \langle \mathcal{H}_I(z_1)\mathcal{H}_I(z_2)\mathcal{O}_0(x)\rangle + \dots\,\,.
\label{eq:Dyson2h}
\end{flalign}
Here, $\mathcal{O}_0(x)=\mathcal{O}[h_0(x),\pi_0(x)]$, and all expectation values of the right-hand side of \eqref{eq:Dyson2h} are taken over the free vacuum.
 The interaction Hamiltonian is determined by~\eqref{eq:Hint}, which we conveniently report here split into cubic and quartic terms
\begin{flalign}
&\mathcal{H}^{(3)}_\text{I}={\lambda v}\left(h^3-3 \langle h^2\rangle h\right)+{\lambda v} (h \pi^2 - \langle \pi^2\rangle h)\,, \label{eq:cubicSF}\\
& \mathcal{H}^{(4)}_\text{I}=\frac{\lambda}{4}\left(h^2+\pi^2-6\langle h^2\rangle-2\langle\pi^2\rangle\right)\left(\pi^2+h^2\right)\,.
\end{flalign}
Here, we truncate the interaction Hamiltonian at fourth order in fluctuations. Higher-order terms only contribute starting at the three-loop level, which lies beyond the scope of this section. Concerning the one-loop chemical potential, we need to evaluate the equation for the chemical potential Eq.~\eqref{eq:radialU1} using the expressions for tree-level correlation functions.
These can be evaluated using the mode decomposition provided in Section~\ref{section:interaction_picture}.\footnote{These read
\begin{equation}
    \langle h_0^2(x)\rangle=\int \frac{{\rm d}^3k}{(2\pi)^3 2} \left(\frac{|Z_+^h|^2}{\omega_+}+\frac{|Z_-^h|^2 }{\omega_-}\right),\quad  \langle \pi^2_0(x)\rangle=\int \frac{{\rm d}^3k}{(2\pi)^3 2} \left(\frac{|Z_+^\pi|^2}{\omega_+}+\frac{|Z_-^\pi|^2 }{\omega_-}\right)\,.
\end{equation}}
One obtains the following one-loop chemical potential:
\begin{flalign}
    \mu_\text{1-loop}^2=m_\text{ph}^2+\lambda_\text{ph} v^2\left\{1+\frac{3\lambda_\text{ph}
    }{4\pi^2}\left[1+f(c_s)\right]\right\}\,,
\end{flalign}
with 
\begin{equation}
   f(c_s)= -\frac{c_s^2}{3}-\frac{\left(1-c_s^2\right)}{3}\sqrt{1+\frac{1}{c_s^2}}+\left(1+\frac{1}{3c_s^{2}}\right)\log \left(\frac{c_s+\sqrt{1+c_s^2}}{\sqrt{1-2 c_s^2}}\right)\,.
\end{equation}
Here $c_s$, as given by Eq.~\eqref{eq:phononspectrum}, reduces to the sound speed of the fluid in the non-relativistic limit. The one-loop chemical potential is time-independent as it is constructed using quadratic correlation functions at the space-time coincidence. Moreover, in the semiclassical limit~\eqref{eq:semiclassicalLimit}, the one-loop correction vanishes, recovering the standard classical expression for the relativistic chemical potential. 

The chemical potential is divergent if expressed in terms of the bare parameters of the theory. However, it is finite once the theory is renormalized. This is possible by introducing the following physical parameters
\begin{equation}
 m^2_\text{ph}=m^2+ \frac{\lambda}{2\pi^2}\left\{\Lambda ^2+\frac{m^2}{2}-m^2 \log \left(\frac{2 \Lambda}{m} \right)\right\}, \qquad \lambda_\text{ph}=\lambda-\frac{5\lambda^2}{4\pi^2}\log\left(\frac{2 \Lambda}{m} \right)\,,
 \label{eq:RenConditionsSF}
\end{equation}
with $\Lambda$ an ultraviolet regulator. 
Notice that all the divergences can be cured by the same counterterms that are needed to renormalize the theory in the vacuum: no additional infinities appear.

This is consistent with the analog derivation performed in Ref.~\cite{Joyce:2022ydd,Nicolis:2023pye}, where the authors evaluated the one-loop corrections to the effective potential in a general $U(1)$ invariant theory, finding that field-independent prescriptions make the effective potential finite.  The same was found in Ref.~\cite{Berezhiani:2021gph} for the one-loop corrections to the dynamics of a real scalar condensate. Furthermore, we reproduce the result of Ref.~\cite{Nicolis:2023pye} that, in the limit of vanishing condensate amplitude $v$, the physical mass~\eqref{eq:RenConditionsSF} matches the pole mass of the vacuum theory.\footnote{It is important to stress that these properties do not hold universally. Not all states have their dynamics regularized by the same counterterms used in the vacuum theory. Discrepancies in this direction often hint at issues with the chosen configuration, such as a divergent energy density in the system. See Refs.~\cite{Baacke:1999ia, Berezhiani:2023uwt,Berezhiani:2021gph}. } 

\subsection{Explicit check of the two-loop stability}
The evaluation of the chemical potential can be systematically extended to arbitrary loop orders. For example, to obtain the two-loop corrected chemical potential, one needs to evaluate Eq.~\eqref{eq:radialU1} by inserting the one-loop corrected quadratic correlation functions $\langle  h^2\rangle$ and $\langle\pi^2\rangle$, and the tree level correlation function $\langle h^3\rangle$ and $\langle h \pi^2\rangle$.

Although computing the full expression is algebraically cumbersome, it is straightforward to show that the resulting correlation functions are time-independent, as long as the time integrals in the Dyson expansion are properly regulated in the far past. This follows from the fact that the integrands depend only on differences of the form $t_i - t$, where $t_i$ are internal times. By redefining time integration variables, the explicit dependence on the external time $t$ can be removed from the integrand and shifted to the lower integration limits. Since the time contour begins at $-\infty_\pm$, shifting the contour by a finite and real amount does not affect the damping behavior at early times. Therefore, no contributions can be obtained from the lower integration extreme, and the external time-label drops out from the computation. To illustrate this, let us consider an explicit example by evaluating the cubic correlator $\langle h^3\rangle$. This yields (schematically)
\begin{flalign}
    \langle h^3\rangle\sim\sum_{\alpha,\beta,\gamma=\pm} \int \rd^3k_1\rd^3k_2 \int_{-\infty_-}^t  \rd t_zf_{\alpha \beta\gamma}(k_1,k_2) e^{-\ri (\omega_{\alpha}(k_1)+\omega_{\beta}(k_2)+\omega_{\gamma}(k_1+k_2)) (t_z-t)}\nonumber\\\sim\sum_{\alpha,\beta,\gamma=\pm}  \int \rd^3k_1\rd^3k_2 \int_{-(\infty+t)+(\ri \epsilon) \infty}^0  \!\! \!\!\!\!\!\! \!\!\!\!\rd\tilde{t}\,f_{\alpha \beta\gamma}(k_1,k_2) e^{-\ri (\omega_{\alpha}(k_1)+\omega_{\beta}(k_2)+\omega_{\gamma}(k_1+k_2)) \tilde{t}}\,,
    \label{eq:cubich}
\end{flalign} 
where we introduced $\tilde{t}=t_i-t$. Here, $f_{\alpha \beta \gamma}$ are the time-independent functions of the spatial momenta $k_1$ and $k_2$, that emerge after Wick contractions are performed.
Hence, since the contribution from the lower limit is exponentially suppressed by the imaginary time rotation, and this damping is unaffected by the shift in integration variables, the contribution from that boundary remains null.
As a result, the correlation functions—and hence the quantum corrections to the chemical potential — remain time-independent. Notice that the time-independence of $f_{\alpha \beta \gamma}(k_1,k_2)$ follows from the time-independence of both the wavefunction coefficients of the quadratic theory and the coefficients in the interaction Hamiltonian. These two are the same properties we exploited in order to demonstrate the property (i) of Section~\ref{sec:state}.

At this stage, we have limited the analysis to equal time and space correlation functions. However, the same result can be extended to correlation functions at arbitrary spatial points. In particular, the key property used in~\eqref{eq:cubich} is that one can always perform a time redefinition to shift the time label 
$t$ to the lower limit of the integration. This property holds regardless of the spatial dependence of the correlator. 

A more interesting and non-trivial equation to examine is Eq~\eqref{eq:angularU1} at order $\hbar^2$. As shown in Section~\ref{sec:state}, this equation is satisfied because equal time correlators involving an odd number of $\pi$ fields vanish, due to the invariance of the kinetic mixing under a combined $\mathbb{Z}_2$ transformation and time reversal — provided the interacting vacuum ensures time-independence of correlators at coincidence.

To check this, we start by studying the one-loop corrections to this equation. They are obtained by evaluating the quadratic correlation function at the leading order in the interaction Hamiltonian, hence the object $\langle \pi_0(x) h_0(x)\rangle$. Since these are needed in the remainder of this section, we provide all tree-level correlation functions
\begin{flalign}
    \langle \pi_0(x) \pi_0(y)\rangle=\int \frac{{\rm d}^3 k}{(2 \pi)^32}\left(\frac{|Z_+^\pi|^2}{\omega_+}e^{-{\rm i} \omega_+ (t-t')}+\frac{|Z_-^\pi|^2 }{\omega_-}e^{-{\rm i} \omega_- (t-t')}\right) e^{{\rm i} \vec{k} \cdot (\vec{x}-\vec{y})}\,,
    \label{eq:SSF2}
    \\
    \langle h_0(x) h_0(y)\rangle=\int \frac{{\rm d}^3 k}{(2 \pi)^32}\left(\frac{|Z_+^h|^2}{\omega_+}e^{-{\rm i} \omega_+ (t-t')}+\frac{|Z_-^h|^2 }{\omega_-}e^{-{\rm i} \omega_- (t-t')}\right) e^{{\rm i} \vec{k} \cdot (\vec{x}-\vec{y})}\,, 
    \label{eq:SSF1}
    \\
     \langle h_0(x) \pi_0(y)\rangle=-\langle  \pi_0(y) h_0(x)\rangle=\int \frac{{\rm d}^3 k}{(2 \pi)^32}\frac{(Z_+^\pi)^*Z_+^h}{\omega_+}\left(e^{-{\rm i} \omega_+ (t-t')}-e^{-{\rm i} \omega_- (t-t')}\right) e^{{\rm i} \vec{k} \cdot (\vec{x}-\vec{y})}\,.
     \label{eq:SSF}
\end{flalign}
In particular, from Eq.~\eqref{eq:SSF}, we see that if we set $t=t'$, the off-diagonal correlation function vanishes, and Eq~\eqref{eq:angularU1} is satisfied at order $\hbar$.

We now turn to two-loop corrections by computing the correlation functions appearing in Eq.\eqref{eq:angularU1} up to order $\hbar^2$.
We start by evaluating the cubic correlation functions, whose first-order correction is determined by the first term of Eq.~\eqref{eq:Dyson2h}
\begin{flalign}
    \langle \pi h^2+\pi^3\rangle=-2\lambda v~\text{Im}\int_{\infty_-}^t {\rm d}^4z ~\big\langle &\left(h_0^3(z)-3 \langle h_0^2\rangle h_0(z) +h_0(z) \pi_0^2(z) -2 \langle \pi_0^2\rangle h_0(z)\right)\nonumber  \\\qquad &\times(\pi_0(x) h_0^2(x)+\pi_0^3(x))\big\rangle\,.
\label{eq:cubiccoindicdence}
\end{flalign}

This contribution vanishes, and the cancellation requires two key observations. First, Eq.~\eqref{eq:cubiccoindicdence} contains an odd number of $\pi$ fields. Therefore, at least one off-diagonal term $\langle h_0(z), \pi_0(x) \rangle$ must appear after performing Wick contractions. In contrast to diagonal contractions, which are proportional to squared norms of the $Z$-factors, off-diagonal ones carry an overall factor of $\mathrm{i}$, as can be seen by inspecting the tree-level correlation functions in Eqs.~\eqref{eq:SSF2}–\eqref{eq:SSF}. This originates from the fact that the product $Z^\pi_+ Z^h_+$ is purely imaginary.

Second, the time integral produces two contributions: one from the lower limit, which is exponentially damped by the $\mathrm{i}\epsilon$ prescription, and one from the upper limit, which gives a factor $\mathrm{i}/(\omega_1 + \omega_2 + \omega_{3})$, where $\omega_i = \omega(k_i)$ denote the frequencies associated with internal momenta $k_i$. At equal times $t = t_1$, the oscillatory terms become unity. As a result, we collect two powers of $\mathrm{i}$ — one from the contraction and one from the integral — yielding an overall real result. Since only the imaginary part contributes to the integral, the entire term vanishes.

This cancellation does not hold for correlators with an even number of $\pi$ fields, where off-diagonal contractions appear in pairs and can yield an imaginary result. Nor does it apply when the fields are evaluated at different times, where the oscillatory prefactors no longer trivialize.

Then, we move to the one-loop corrections to the quadratic correlation function $\langle h\pi\rangle$. The tree-level term is vanishing as we mentioned (see Eq.~\eqref{eq:SSF}), and we have to evaluate both the first and second-order corrections in the interaction Hamiltonian, with the first-order correction involving $\mathcal{H
}_I^{(4)}$ and the second-order $\mathcal{H}_I^{(3)}$.
 The first-order correction vanishes for the same reason as the correction to the cubic correlation functions. Concerning the second-order corrections, these are also vanishing. In particular, the last integral of Eq.~\eqref{eq:Dyson2h} is imaginary due to the additional time-integral, that generates an additional power of $\rm i$ compared to the previous order. Therefore, the real part of the integral is null. We are left with the third term of Eq.~\eqref{eq:Dyson2h}. 
To see that this term is also vanishing, we write it as
\begin{flalign}
     \int_{-\infty}^{t} \rd^4z_1 \int_{-\infty}^{t} & \rd^4z_2 \langle \mathcal{H}_I({z_1)} h_0(x) \pi_0(x) \mathcal{H}_I({z_2)} \rangle\nonumber\\&= 2\text{Re}  \int_{-\infty}^{t} \rd^4z_1 \int_{-\infty}^{t_1} \rd^4z_2 \langle \mathcal{H}_I({z_1)} h_0(x) \pi_0(x) \mathcal{H}_I({z_2)} \rangle\,,
\end{flalign}
and the same conclusion applies.
We emphasize that, in reshuffling the integration ranges, we neglect the difference in the imaginary time rotations of the two time integrals. These ${\rm i}\epsilon$ prescriptions are solely responsible for damping contributions in the far past and do not affect the evaluation of finite corrections.

Therefore, correlation functions entering Eq.~\eqref{eq:angularU1} vanish individually up to $\hbar^2$ corrections, which is a check of the all-order proof of Section~\ref{sec:state}.

\subsection{Beyond equilibrium: Gaussian states and unstable superfluids}
\label{eq:UnstableSF}

The property that provides a stable superfluid is the time independence of equal-time correlation functions of fluctuations in our parameterization. The interacting vacuum for fluctuations provides a state for which this property is satisfied.
However, in general, not all states whose classical dynamics reproduce the classical solution~\eqref{classical_superfluid} retain this property once quantum corrections are taken into account. In particular, different states could share the same (semi-)classical limit, but could have a completely different quantum evolution. A prominent example is given by coherent states, which we show now can approximate a stable superfluid configuration up to order $\hbar$, but ultimately fail to remain stationary in the full quantum theory.

 The example we consider in this section is the free vacuum for fluctuations:
\begin{equation}
|\tilde{v}\rangle = |0_{h,\pi}\rangle = e^{-{\rm i} f} e^{-{\rm i} S} |0\rangle\,,
\end{equation}
 which can be represented as a squeezed coherent state built on top of the free fundamental vacuum of the theory, as we mentioned in App.~\ref{App:squeezing}.
 
It was observed in~\cite{Berezhiani:2023uwt} that coherent or squeezed states—such as the above free vacuum —  develop pathologies starting from two-loop corrections to the background, such as a divergent energy density of the system. This stems from the fact that non-Gaussian components are essential to define a physically consistent configuration in an interacting quantum field theory. This dressing, however, still leaves room to add physical non-Gaussian components to the state.
We now show that without the specific non-Gaussian dressing — implemented in~\eqref{eq:non-gaussianV} by the evolution operator $U_I(t,-\infty_-)$ — the state would have a non-trivial quantum time evolution. 

To study the time evolution of the background depicted by the free vacuum, we generalize the framework of the previous section to more general time-dependent configurations. Therefore, we introduce the following general field reparametrization
\begin{equation}
\Phi(x)=\frac{1}{\sqrt{2}}\left(v(t) + h(x) + \ri \pi(x)\right)e^{\ri \mu(t) t}\,,
\label{eq:SFdecomposition2}
\end{equation}
with the following tadpole conditions imposed on $h(x)$ and $\pi(x)$:
\begin{equation}
    \langle \tilde{v}|h(x)|\tilde{v}\rangle=  \langle \tilde{v}|\pi(x)|\tilde{v}\rangle=0 \qquad\Longrightarrow \qquad \langle \tilde{v}|\Phi(x)|\tilde{v}\rangle=\frac{1}{\sqrt{2}}v(t)e^{\ri \mu (t) t}\,.
\end{equation}
At the classical level (so ignoring fluctuation fields), this configuration describes a general time-dependent configuration with a net charge density 
\begin{equation}
   n_\text{cl}(t)=v_\text{cl}^2(t)\frac{d}{dt}(\mu_\text{cl} t)\,.
    \label{eq:classicalcharge}
\end{equation}
This charge density is classically conserved, in the sense that its time derivative vanishes upon imposing the classical equations of motion. In particular, the imaginary part of the equation of motion evaluated on the classical configuration yields precisely the constraint $\dot{Q}_\text{cl}=0$.

Now, let us derive the analog of Eq.~\eqref{eq:radialU1} and Eq.~\eqref{eq:angularU1}. We find
\begin{flalign}
(\mu+\dot{\mu}t)^2-\frac{\ddot{v}}{v}=m^2 +\lambda v^2+3\lambda \langle h^2\rangle+\lambda   \langle \pi^2\rangle+\frac{\lambda}{v}\langle h^3\rangle+ \frac{\lambda}{v}\langle h \pi^2\rangle\,,
\label{eq:radialU1td}\\
\frac{\partial}{\partial t}n(t)+2\lambda v^2\langle h\pi\rangle+{\lambda }v\langle \pi h^2+\pi^3\rangle=0\,,
\label{eq:angularU1td}
\end{flalign}
where expectation values are now considered over the $|\tilde{v}\rangle$ state.
Eq.~\eqref{eq:radialU1td} generalizes equation~\eqref{eq:radialU1} by introducing an additional term proportional to the acceleration of the amplitude. However, the most interesting generalization is provided by Eq.~\eqref{eq:angularU1td}, which now contains an additional contribution proportional to the time derivative of the function $n(t)=v^2(t)\frac{d}{dt}(\mu t) $. 

We refer to this quantity as the \textit{background charge density}—
the analog of ~\eqref{eq:classicalcharge}, but constructed from the one-point function of the field operator. For translationally invariant states, charge densities are trivially connected to charges by a volume factor at all times. 

Once quantum corrections are included, the background charge is generally no longer conserved in contrast to its classical counterpart: its time evolution is sourced by correlation functions of the fluctuation fields. In the full quantum theory, it is the expectation value of the charge operator over the full quantum state that remains conserved. This implies the following identity that involves the background charge
:
\begin{equation}
    \partial_t \langle {Q} \rangle=\dot{Q}+\dot{Q}_\text{fluct}=0,
    \label{eq:charge}
\end{equation}
with 
\begin{equation}
  Q_\text{fluct}\equiv \int \rd^3 x\,\left(  \langle\Pi^\pi(x)h(x)\rangle-\langle\Pi^h(x)\pi(x)\rangle\right)\,,
\end{equation}
where we introduced a second type of charge, the $\textit{fluctuation charge}$. The above equation is obtained by considering the definition of a $U(1)$ charge, and then applying the field redefinition~\eqref{eq:SFdecomposition2}. 

We now provide the physical interpretation of the two types of charges introduced above. The background charge corresponds to the charge stored in the condensate and represents the number density of particles at zero momentum. In contrast, the fluctuation charge accounts for the charge carried by excitations propagating on top of the condensate, which describe $\Phi$-particles with finite momentum. For the superfluid to remain stable, these two components must be independently conserved, as expressed by Eqs.~\eqref{eq:angularU1td} and~\eqref{eq:charge}. This condition is enforced, at the technical level, by the vanishing of equal-time correlation functions that are odd in $\pi$, a property satisfied by the interacting vacuum for fluctuations, for example. 

Unlike the conservation of the total charge, which follows from the $U(1)$ symmetry of the theory, the separate conservation of background and fluctuation charges is not guaranteed in general. It is enforced only for specific states, such as the interacting vacuum for fluctuations. In more general configurations, fluctuations can source a non-trivial time evolution, leading to charge exchange between the condensate and its excitations. In such cases, the background charge evolves precisely to compensate for the time dependence introduced by the fluctuation sector. For example, introducing a non-trivial fluctuation mode $a_\pm(k)$ on top of the vacuum can lead to rescattering with the zero-momentum particles in the condensate. This interaction transfers momentum to the background, thereby altering the configuration of the zero-momentum component, leading to a non-trivial time-evolution. In the phonon language, this process corresponds to the decay of a phonon into multiple phonons.

We show now that the free vacuum for fluctuations $|0_{h,\pi}\rangle$ provides a class of such unstable configurations. We evaluate correlation functions over $|\tilde{v}\rangle$ entering~\eqref{eq:radialU1td} and~\eqref{eq:angularU1td}. Using the decomposition~\eqref{eq:ExpFl} and setting $t_*=0$, correlators over the free vacuum can be decomposed as
\begin{flalign}
    \langle \tilde{v}|\mathcal{O}(x)|\tilde{v}\rangle=& \langle 0_{h,\pi}|U^\dagger(0,t) \mathcal{O}_0(x) ~ U(0,t)|0_{h,\pi}\rangle\simeq \langle \mathcal{O}_0(x)\rangle
- 2\,\text{Im} \int_{0}^t  \rd^4 z_1\, \langle \mathcal{H}_I(z_1)\mathcal{O}_0(x)\rangle+\ldots \,\,.
\label{eq:dysonOOE}
\end{flalign}
In contrast to the expansion~\eqref{eq:Dyson2h}, here the early-time boundary of the evolution operator is $t_z=0$, and not $t_z=-\infty_-$. However, this difference is important only for loop corrections to quadratic correlation functions, and for tree-level contributions to cubic (or higher-order) correlators. All these corrections arise at order $\hbar^2$ or higher, indicating that deviations from stationarity are expected to appear at this order when the free vacuum is used.

This also justifies why we do not rederive the interaction picture decomposition or modify the fluctuation equations to include explicit time dependence in $\mu(t)$ and $v(t)$. Incorporating such time dependences would require redefining the wavefunction coefficients or reorganizing the formalism by including time-dependent coefficients in the interacting Hamiltonian. However, since we are focused on the leading deviation, all relevant correlators can be computed using the same decomposition as in Section~\ref{section:interaction_picture}, with the only change being the time contour. Corrections from time dependences of $\mu$ and $v$ start being important at higher-loop orders.
 
We now show that correlation functions evaluated on the state $|\tilde{v}\rangle$ that are odd in $\pi$ do not vanish at equal times and are, in general, time-dependent. These features imply that Eqs.~\eqref{eq:radialU1td} and~\eqref{eq:angularU1td} cannot be satisfied by a constant background charge and chemical potential.
We do this by focusing on the cubic correlation function. Using Eq.~\eqref{eq:dysonOOE}, the first non-trivial corrections read (schematically):
\begin{flalign}
\langle \tilde{v}| \langle \pi h^2+\pi^3\rangle|\tilde{v}\rangle=& -2\,\text{Im}\int_{0}^t  {\rm d}^4z \, \langle \mathcal{H}^{(3)}_\text{I}(z) ( \pi_0(x) h_0^2(x)+\pi_0^3(x))\rangle \nonumber \\=
&\sum_{\alpha,\beta,\gamma=\pm}\int {\rm d}^3 k_1 {\rm d}^3k_2\,  g_{\alpha\beta\gamma}(k_1,k_2) \sin [(\omega_\alpha(k_1)+\omega_\beta(k_2)+\omega_\gamma(k_1+k_2))t]\,,
 \label{eq:gaussianDysonSeries}
\end{flalign}
which is non-vanishing for $t>0$. The functions $g_{\alpha \beta \gamma}(k_1, k_2)$ arise from the momentum dependence of the field decompositions, analogously to what occurs in Eq.~\eqref{eq:cubich}. These are combinations of the wave-function coefficients and of the positive and negative frequencies.

We performed the integral over the internal time. Since the integration starts at  $t_z = 0$, there is no exponential damping at the lower limit, and a time-dependent contribution arises from this specific boundary. More generally, this shows that correlation functions evaluated over $ |\tilde{v}\rangle $ exhibit time dependence even when all operators are inserted at equal times.
Nonetheless, the cubic correlation function vanishes exactly at $ t = 0 $. This follows from the fact that $ |\tilde{v}\rangle $ was constructed to be a Gaussian state at $ t = 0 $, for which the connected part of correlation functions beyond the quadratic order vanishes.

This class of time-dependencies makes the time-derivative of the background charge non-zero, indicating a transfer of the charge density from the one-point function to higher-order correlation functions. Of course, one should rigorously evaluate the full effect and verify that the different time dependencies do not cancel out. However, this issue has already been addressed in the numerical analysis of Ref.~\cite{Kovtun:2020ndc, Kovtun:2020udn}, where the authors investigated the dynamics of a complex scalar field in 1+1 dimensions, initialized on the state $|\tilde{v}\rangle$. There, it was shown that initializing correlation functions with the free vacuum leads to a time-dependent one-point function for the condensate, starting from the two-loop order\footnote{As previously discussed, this approach breaks down beyond leading order due to its failure to regularize key observables. Nonetheless, it remains informative for probing certain aspects of the background evolution, as demonstrated in\cite{Berezhiani:2021gph} in the case of a real scalar condensate.}. 

As mentioned, we can interpret the transfer of charge in terms of particle content. The state $|v\rangle$ represents a condensate of fundamental particles, all at zero momentum and of the same global charge. Because of this, unless an external perturbation is introduced, the system evolves trivially: no number-changing process can take place due to the conserved charge and no scattering among constituents can occur for kinematic reasons. In contrast, the state $|\tilde{v}\rangle$ is interpreted as a collection of particles in which a part resides in the condensate and a small part in fluctuations. The time evolution of this system is non-trivial since the rescattering of excited particles on the constituents can move part of the charge from the background to the excited component, triggering the instability.

\section{One-loop gaplessness of the Goldstone}
\label{sec:gaprel}

We conclude by checking that the one-loop spectrum of superfluid perturbations contains a gapless Goldstone mode. This provides a perturbative check of the validity of the Goldstone theorem beyond the semiclassical limit~\eqref{eq:semiclassicalLimit}, where loop corrections cannot be neglected anymore.

The existence of a gapless excitation is a robust, non-perturbative consequence of spontaneous symmetry breaking. It holds regardless of whether the state that breaks the internal symmetry is Lorentz invariant~\cite{Goldstone:1962es} or not~\cite{Nielsen:1975hm}. 
Nonetheless, several perturbative computations suggest an apparent violation of this property for the superfluid spectrum once loop corrections are included.
These analyses are typically performed using the two-particle-irreducible (2PI) formalism~\cite{PhysRevD.10.2428, Berges:2004yj} within the Hartree approximation (see e.g.\cite{RevModPhys.75.121}). Although most of these studies are carried out at finite temperature, the apparent emergence of a gap persists even in the zero-temperature limit \cite{Alford:2013koa, Sharma:2018ydn, Sharma:2022jio, Pilaftsis:2013xna}. 
This issue can be circumvented by explicitly imposing the gaplessness, as achieved in \cite{Sharma:2018ydn, Sharma:2022jio}
 by the introduction of a second chemical potential for fluctuations.
We demonstrate that, within our approach, the gaplessness of the Goldstone mode is preserved naturally at one loop, in full agreement with the non-perturbative theorems.

In the context of the 2PI formalism, several solutions have been proposed to address the emergence of the gap. These rely on ad-hoc modifications of the generating functional \cite{Ivanov:2005yj,Pilaftsis:2013xna, YUKALOV2008461}, and they all give up the self-consistency of the computation. 
Here, we show that, in the context of relativistic superfluids, quantum corrections do preserve the gaplessness of the Goldstone if vacuum polarization diagrams are properly accounted for. 
These diagrams are usually discarded in the Hartree approximation, which works by neglecting the setting sun diagram in the 2PI effective action while retaining the double bubble diagram. 

 The decomposition of fluctuations that makes the gaplessness of the phonon manifest corresponds to the perturbation of the modulus and the phase of the scalar field $\Phi\propto(v+\tilde{h})e^{\ri(\mu t+\tilde{\pi})}$, rather than the linear decomposition $\Phi\propto(v+h+\ri\pi)e^{\ri\mu t}$. The former parametrization takes advantage of the fact that the spontaneously broken $U(1)$ is realized as a shift symmetry on $\tilde{\pi}$, while the latter one obscures the nonlinearly realized symmetry. However, the linear decomposition emerges more naturally in our implementation of the background field method for coherent states. Moreover, while perfectly legitimate in the case of spontaneous symmetry breaking in the vacuum, a concern has been raised in the literature regarding the applicability of the $\tilde{\pi}$-parameterization in the presence of nonzero charge density \cite{Rajaraman:1975qr}. As such, we focus on demonstrating the cancellation of spurious contributions of the Hartree approximation to the phonon mass-gap within the linear parameterization.

\subsection{Equations of motion for quadratic correlation functions}

In this section, the analysis of correlation functions is performed in $(\vec{q}, t)$ representation.
To verify the presence of a gapless mode, we show that there is a one-loop corrected quadratic correlation function that exhibits a pole at $|\vec{q}|=0$ in momentum space. 

For convenience, we drop the vector symbol on momenta in the reminder of this section.

To begin with, we consider the operatorial equations of motion \eqref{eq:radialEqCondensate} and \eqref{eq:angularEqCondensate}, and we convert them into equations for the quadratic correlation functions by multiplying by $h(y)$ and $\pi(y)$, and taking the expectation value over the unperturbed superfluid state $|v\rangle$.  This provides a system of \textit{four} equations of motion for the quadratic correlation functions, with the diagonal ones reading
\begin{flalign}
\biggr\{\Box_x-3\lambda\langle h^2\rangle-\lambda\langle \pi^2\rangle\biggr\}\langle \pi(y)\pi(x)\rangle+&2\mu\partial_t\langle \pi(y){h}(x)\rangle+2\lambda v\langle \pi(y)\pi(x) h(x) \rangle\nonumber+\\+\lambda v\langle \pi(y)\pi(x)^2\rangle+\lambda& \langle \pi(y) \pi(x)^3\rangle+ \lambda\langle \pi(y) \pi(x) h(x)^2\rangle+\ldots=0\,,
\label{eq:angular2P}
\end{flalign}
and
\begin{flalign}
\biggr\{\Box_x+2\lambda v^2-3\lambda\langle h^2\rangle-&\lambda\langle \pi^2\rangle\biggr\} \langle h(y)h(x)\rangle\nonumber -2\mu\partial_t\langle h(y){\pi}(x)\rangle+3\lambda v \langle h(y)h(x)^2\rangle+\\ &+\lambda v\langle h(y)\pi(x)^2\rangle +\lambda \langle h(y) h(x)^3\rangle+\lambda\langle h(y) h(x) \pi(x)^2\rangle+\ldots=0\,.
\label{eq:EOM2P}
\end{flalign}

 To keep the expression compact, we made a truncation up to the quartic order in fluctuations, since these are the relevant terms for the loop order we are working at.
Off-diagonal equations are obtained from the one above with the replacements
\begin{equation}
    h(y)\to \pi(y) \quad \text{and} \quad   \pi(y)\to h(y)
\end{equation}
in the first and second equation, respectively.

In what follows, we consider the momentum representation of the set of equations above. 
We introduce the following notation for the momentum representation of correlators:
\begin{equation}
    \langle \mathcal{O}(x,y)
    \rangle_q =\int \rd^3 \vec{r}\, \langle \mathcal{O}(x,y) \rangle e^{-{\rm i} {q}\cdot (\vec{x}-\vec{y})}\,.
\end{equation}
The subscript $q$ indicates that the Fourier transformed correlation function is a function of the external momentum $q$.

At the order we are working, we can perform Wick contractions of the quartic correlation functions to reduce them to the products of quadratic correlation functions. We obtain:
\begin{flalign}
&\biggr\{\partial_t^2+ q^2+2\lambda\left(\langle \pi_0^2\rangle-\langle h_0^2\rangle\right)\biggr\}\langle \pi(y)\pi(x)\rangle_q+2\mu\partial_t\langle \pi(y){h}(x)\rangle_q+2\lambda v\langle \pi(y)\pi(x) h(x) \rangle_q=0\,,
\label{eq:angular2PF}\\
&\biggr\{\partial^2_t+ q^2+2\lambda v^2 \biggr\}\langle h(y) h(x)\rangle_q  -2\mu\partial_t\langle h(y){\pi}(x)\rangle_q+\lambda v \left(3\langle h(y)h(x)^2\rangle_q+\langle h(y)\pi(x)^2\rangle_q \right)=0\,.
\label{eq:radial2P}
\end{flalign}
The equations of motion for the non-diagonal correlation functions are obtained by replacing $\pi(y)$ with $h(x)$ in the above equations, and vice versa. These read
\begin{flalign}
&\biggr\{\partial_t^2+ q^2+2\lambda\left(\langle \pi_0^2\rangle-\langle h_0^2\rangle\right)\biggr\}\langle h(y)\pi(x)\rangle_q+2\mu\partial_t\langle h(y){h}(x)\rangle_q+2\lambda v\langle h(y)\pi(x) h(x) \rangle_q=0\,, \label{eq:angular2off}
 \\
&\biggr\{\partial^2_t+ q^2 +2\lambda v^2 \biggr\}\langle \pi(y) h(x)\rangle_q  -2\mu \partial_t\langle \pi(y){\pi}(x)\rangle_q+\lambda v \left(3\langle \pi(y)h(x)^2\rangle_q+\langle \pi(y)\pi(x)^2\rangle_q \right)=0\,.
\label{eq:radial2Poff}
\end{flalign}

Notice that we have promoted the correlation function $\langle \pi_0(y) \pi_0(x)\rangle$ originating from the contraction of quartic diagrams to the full interacting correlator $\langle \pi(y) \pi(x)\rangle$. This is justified, as they start differing only at higher orders in the loop expansion. The advantage is that, in this form, one can easily read that quartic correlation functions provide an effective gap for the Goldstone mode.

\subsection{Quartic correlation functions and the anomalous average}
We start by evaluating the correlation function $\langle \pi_0^2-h_0^2\rangle$,
which enters in Eqs.~\eqref{eq:angular2PF} and~\eqref{eq:angular2off}. Notice that, if written in terms of the fundamental field of the theory, it originates from the class of correlators
\begin{equation}
   \langle 0_{h,\pi} | {\pi}_0^2(x)- {h}_0^{2}(x)|0_{h,\pi}\rangle \sim  e^{-2{\rm i}\mu t}\langle v|  {\Phi}(x) {\Phi}(x)|v\rangle \,,
   \label{eq:quartic}
\end{equation}
which would vanish if we replace the superfluid state with the fundamental $U(1)$-invariant vacuum of the theory.

This correlation function can be evaluated by performing Wick contractions and using the ladder expansion for the real and imaginary fields:
\begin{flalign}
  \langle \pi_0^2-h_0^2\rangle=&\nonumber\int \frac{{\rm d}^3k}{(2\pi)^3 2} \left(\frac{|Z_+^\pi|^2-|Z_+^h|^2}{\omega_+}+\frac{|Z_-^\pi|^2 -|Z_-^h|^2 }{\omega_-}\right)\\
   = &\frac{\mu^2 c_s^2}{2 \pi^2 }\left\{c_s^2\left(1-\sqrt{1+c_s^{-2}}\right)+\log\left(\frac{2\Lambda}{\mu}\right)-\log\left({\sqrt{1+c_s^{2}}+c_s}\right)\right\}\,.
   \label{eq:anomalous}
\end{flalign}
This is a fully relativistic result and it is divergent. In particular, by employing cutoff regularization with a UV regulator $\Lambda$, this term appears to be logarithmically divergent.  
We can also appreciate that the result is vanishing if we set $v= 0$.

If we were to neglect cubic correlation functions from the system of equations \eqref{eq:quartic}, we would find that this term contributes to the Goldstone dispersion relation by introducing a logarithmically divergent gap term. Unless offset by other contributions, this divergence poses a significant issue, as renormalizing it away would require a counterterm that explicitly breaks the $U(1)$ invariance of the fundamental Lagrangian \eqref{eq:LagrSF}. However, as we show now, cubic correlation functions come into play and are essential both in curing the divergence and in restoring the gaplessness of the Goldstone at one-loop. 

We complete this part by mentioning that, when the non-relativistic limit of the theory is considered and we reduce to the theory of a single degree of freedom $a_{-}$, the term~\eqref{eq:quartic} becomes proportional to the zero temperature limit of the so-called \textit{anomalous average} $\sigma_{k} \sim \langle v|(a_-)_{{k}}(a_-)_{-{k}}|v\rangle$, albeit integrated over all momenta $k$. The above correlation function is non-zero only because the finite charge density of the state $|v\rangle$ spontaneously breaks the $U(1)$ global invariance of the theory. In the case of non-relativistic superfluids, the integrated anomalous average generates the gap to the Goldstone in the Hartree approximation \cite{Sharma:2018ydn}.

\subsection{Evaluating cubic correlation functions in the diagonal equations of motion}

We now move to cubic correlation functions that enter in the equations of motion \eqref{eq:angular2PF}. These are the terms that would be missed in the 2PI analysis if the setting sun diagram were neglected in the effective action.

Given the nontrivial background, it is not possible to evaluate them exactly at one-loop. Nevertheless, we can study their momentum representation by expanding the integrand in a series of small external momenta $q$.
As demonstrated in Appendix \ref{app:correlationfunction}, we have
\begin{equation}
\lim_{q/\mu \to 0} \, \langle \pi(y) \pi(x) h(x)\rangle_q= \delta m_\pi^2 \langle \pi(y) \pi(x)\rangle_q+\dots\,\,.
\label{eq:cubicPi}
\end{equation} 
Here, we isolated the contribution that scales as $q^{-1}$, which provides a gap for the Goldstone as in the previous section. Ellipses represent terms that approach a constant in the limit $q\to 0$. 
Notice that these residual corrections are \textit{UV-finite} and the only UV-sensitive contribution of this correlation function is provided by the first term on the right-hand side of \eqref{eq:cubicPi}.
The coefficient $\delta m_\pi^2 $ is given by the following expression
\begin{flalign}
\delta m_\pi^2 =&-{\lambda v}\sum_{\beta=\pm}\int \frac{{\rm d}^3k}{2(2 \pi)^3}\left\{\frac{|Z^h_\beta|^2|Z^\pi_\beta|^2}{\omega_\beta^3 }\right.\left.+\frac{|Z_+^h|^2|Z_-^\pi|^2+|Z_-^h|^2|Z_+^\pi|^2}{\omega_-\omega_+(\omega_-+\omega_+)}+\frac{|Z^h_\beta|^2|Z^\pi_\beta|^2(\omega_+-\omega_-)^2}{2\omega_\beta^2\omega_+\omega_-(\omega_-+\omega_+)}\right\}\nonumber \\
&=-\frac{1}{v}\langle \pi_0^2-h_0^2\rangle\,.
\label{eq:Beta1}
\end{flalign}
To keep the expressions of these factors manageable, we suppressed the $k$-dependence in $\omega_\pm$ and in the wavefunction factors $Z_\alpha^{h,\pi}$ that enter \eqref{eq:Beta1}. We can check the identity of $\delta m^2_\pi$ with the correlation function \eqref{eq:quartic} by explicitly evaluating the integral. As we can see, the cubic correlation function matches exactly the quartic contribution \eqref{eq:quartic}, canceling the divergence and removing the mass term in Eq~\eqref{eq:angular2off}.

We also evaluate the divergent part of the equation of motion for $\langle h(y) h(x)\rangle$, which requires evaluating the cubic correlation function $\langle 3 h(y) h(x)^2+h(y)\pi(x)^2\rangle$ in Eq.~\eqref{eq:angular2off}. In this case, this contribution is not diverging as $q\to 0$. Nevertheless, we get divergences from integrating over loops. In particular, the divergent part is provided by the contribution (see again Appendix \ref{app:correlationfunction})
\begin{equation}
    \langle 3 h(y) h(x)^2+h(y)\pi(x)^2\rangle_q=\left(\delta m^2_h\right)_\infty \langle h_0(y) h_0(x)\rangle_q+\ldots\,\,.
\end{equation}
with
\begin{flalign}
   \left(\delta m^2_h\right)_\infty=-\frac{5\lambda v}{2\pi^2}\log\left(\frac{2\Lambda}{m_h}\right)\,.
    \label{eq:Beta3}
\end{flalign}
Here, ellipses represent finite contributions, while $\Lambda$ is the UV-cutoff. 
We stress that this contribution does not rely on a low-momentum approximation and provides the full one-loop divergent part of the equation of motion for the real field.

\subsection{Evaluating cubic correlation functions in the off-diagonal equations of motion}

Finally, we evaluate the cubic correlation functions of the non-diagonal equations \eqref{eq:angular2off} and \eqref{eq:radial2Poff}. Similarly to the previous section, we need to evaluate the low-$q$ behavior of the cubic correlation functions of Eq.~\eqref{eq:angular2off} to find the one-loop corrections to the Goldstone propagator,  while we need the divergent part of \eqref{eq:radial2Poff} to show the finiteness of correlation functions after renormalization.  

The  low-momentum limit of the cubic correlation function in Eq.~\eqref{eq:angular2off}, reads
\begin{equation}
   \lim_{q/\mu\to 0} \langle h(y)\pi(x) h(x) \rangle_q=\delta m_\pi^2 \langle h_0(y) \pi_0(x)\rangle_q+ \ldots\,\,.
   \label{eq:cubicOffPI}
\end{equation}

We isolated the contribution proportional to $\delta m_\pi^2$, which is the only UV divergence appearing from this correlation function, as above. Notice that there are no $q^{-1}$ contributions arising from this correlation function. Again, ellipses are
\textit{constant and finite} corrections in ${q}$ to the correlation function. 
Concerning the other off-diagonal correlators Eq.~\eqref{eq:radial2Poff} we find
\begin{equation}
 3\langle \pi(y)h(x)^2\rangle_q+\langle \pi(y)\pi(x)^2\rangle_q=\left(\delta m^2_h\right)_\infty \langle \pi_0(y) h_0(x)\rangle_q +\ldots\, ,
\end{equation}
with $\left(\delta m^2_h\right)_\infty $ given by   \eqref{eq:Beta3} and with ellipses representing finite contributions. As above, the divergent part does not rely on any small $q$ approximation.\footnote{Notice that in this case there seems to be a contraction that could naively generate a term proportional to $\langle \pi_0(y)\pi_0(x)\rangle$ in the $q\to 0$ limit, which would diverge as $q^{-1}$. However, this term drops out once the contribution from the second and third terms of \eqref{eq:DysonSeries} are summed up, see Appendix \ref{app:correlationfunction}}

\subsection{One-loop finiteness of the spectrum and gaplessness of the Goldstone }
\label{sec:gapless}

In the previous sections, we have proven the cancellation of the anomalous average term (usually arising in the Hartree approximation) via the cubic correlation functions.  This cancellation is sufficient to ensure the gaplessness of the phonon mode. We now conclude by examining how the remaining subleading corrections influence the propagator.

In the small momentum limit, we can use our previous results to rearrange the system of equations in the following matrix form:
\begin{equation}
    \begin{pmatrix}
    &\partial_t^2 &   2 {\rm i} \mu \partial_t     \\
   & -2 {\rm i} \mu \partial_t & \partial_t^2+2 \lambda_\text{ph} v^2
\end{pmatrix} \langle \boldsymbol{G}\rangle _q=\boldsymbol{J}_q(t-t'),
\end{equation}
where the matrix
\begin{equation}
    \langle \boldsymbol{G}\rangle_q= \begin{pmatrix}
    &\langle \pi(y)\pi(x)\rangle_q &  \langle h(y) \pi(x)\rangle_q     \\
   &  \langle  \pi(y)h(x)\rangle_q   & \langle h(y)h(x)\rangle_q 
\end{pmatrix} 
\end{equation}
contains the one-loop two-point functions built from the real and imaginary fields.

The matrix $\boldsymbol{J}_q$ on the right-hand side collects all terms that approach a finite constant as $q \to 0$, previously denoted by ellipses in earlier sections. This matrix can be parametrized as follows:
\begin{equation}
J^{ab}_q=\sum_{\alpha=\pm} \sigma^{ab}_\alpha e^{-{\rm i} \omega_\alpha(t'-t)}. 
\label{eq:Jq}
\end{equation}

First, we note that the system of equations is finite: the left-hand side depends only on physical parameters, while the right-hand side involves finite coefficients. 
Indeed, the cancellation of the anomalous average term is important also for the renormalizability of the theory, since it removes the logarithmic divergence in Eq.~\eqref{eq:anomalous} without the need of adding $U(1)$ breaking counterterms in the fundamental Lagrangian.

In contrast, in the equations for $\langle \pi(y) h(x)\rangle$ and $\langle h(y) h(x)\rangle$, the divergent contribution from cubic correlation functions does not cancel. However, it is renormalized by a shift of the coupling constant, coming from the vacuum counterterm \eqref{eq:RenConditionsSF}. This explains the appearance of the physical coupling constant in the above system. We remark once again that the usual renormalization in the vacuum sector is enough to make both the one-point function of the superfluid and the two-point function of its fluctuation fields finite, even if q-corrections are reintroduced.
Note that wavefunction renormalization is not required at this order, consistent with the fact that it is also unnecessary in the vacuum sector at one loop.

After demonstrating the cancellation of the superficial mass term for the Goldstone and renormalizing the fluctuation equations, we discuss the role of the terms that remain finite in the $q \to 0$ limit. 
We solve the system by convoluting the sources with retarded Green's functions, and expand in series for small momentum $q$. We find~\footnote{We remind that Green's functions can be constructed starting from the free theory quadratic correlation function as $G_C=\theta(t_y-t_x)\langle[\phi(y),\phi(x)]\rangle$.}
\begin{flalign}
       \langle \pi(x)\pi(y)\rangle_\text{q}=& \langle \pi_0(x) \pi_0(y)\rangle_q+ \sum_{\alpha=\pm}\int_{-\infty_-}^t {\rm d} t_z \left\{\sigma^{11}_\alpha \langle [\pi_0(y), \pi_0(z)]\rangle_q +\sigma_\alpha^{21} \langle [h_0(y), \pi_0(z)]\rangle_q\right\}e^{-{\rm i}\omega_\alpha(t'-t_z)}\nonumber\\= &\frac{1}{2(2\pi)^3}\frac{|Z_-^\pi|^2}{\omega_-}\left(1-\frac{ \sigma^{11}_-}{2\omega_-}e^{-{\rm i}\omega_- (t'-t)}+\dots\right)e^{-{\rm i}\omega_- (t'-t)} 
       +\ldots\,\,.
\label{eq:1loopPP}  \end{flalign}
Ellipses are loop corrections scaling at most as $q^{-1}$. Moreover, the first term on the left-hand side has the form of a tree-level correlator, albeit with renormalized parameters.

From Eq.~\eqref{eq:1loopPP}, we observe the appearance of a leftover term of order $q^{-2}$.  
While this may seem an awkward contribution, this term is not associated with the appearance of a gap. Indeed, the emergence of a mass gap from the expansion around a massless theory arises from contributions that scale $q^{-3}$.\footnote{If one considers the mass of a scalar field as a perturbation around the massless theory, one would find the following expansion of the propagator
\begin{equation}
   \langle \phi(\vec{x},0)\phi(\vec{y},0)\rangle= \int\frac{\rd^3q}{2q}\left(1+\frac{m^2}{q^2}+\ldots\right)e^{\ri \vec{q}\cdot(\vec{x}-\vec{y})}\sim \int\frac{\rd^3q}{2\sqrt{q^2+m^2}}e^{\ri \vec{q}\cdot(\vec{x}-\vec{y})}\,.
\end{equation}}
This is exactly the behavior we would obtain if the anomalous average were retained.
On the contrary, this term, if non-zero, could be interpreted as a q-correction to the leading phonon dispersion relation, which is beyond the approximation employed. We leave the exact derivation of this contribution for future work.

\section{Outlook}\label{sec:outlook}
In this work, we have pushed the stability analysis to the finish line and demonstrated that for a massive complex scalar field, there exists a special, modified coherent state that corresponds to the homogeneous classical time-dependent background of the superfluid, which is capable of retaining classicality to all orders in perturbation theory. We showed that the time-dependence of correlation functions in the state in question 
fully tracks the
classical dynamics. 
In other words, we have proven that the homogeneous superfluid corresponds, to all orders, to what was named in \cite{Nicolis:2011pv} a Spontaneous Symmetry Probing state. We clarified how the correct choice of non-Gaussianities in the superfluid state is necessary to guarantee the stability of the background. Finally, we have shown that the spectrum of fluctuations consistently preserves a gapless Goldstone mode at one-loop order, and we have identified and resolved certain misconceptions in the literature arising from inconsistent truncations in the treatment of loop corrections to the phonon propagator. 

To conclude, we would like to point out another important aspect of the aforementioned stability feature, namely that it holds even when the initial field amplitude is large enough that the nonlinear part of the potential dominates over the mass term. This regime corresponds to a superfluid with a relativistic sound speed. The possible importance of this feature for effective field theories relevant to cosmology can be conveniently outlined in the example of a linear sigma model
\beq
\mathcal{L}=|\partial_\mu\Phi|^2-\frac{\lambda}{2}\left(|\Phi|^2-v_0^2\right)^2\,.
\eeq
As it is well known, the vacua of this theory correspond to a non-vanishing expectation value of the field, which breaks the $U(1)$ symmetry spontaneously. The superfluid state can be constructed in this theory too \cite{Nicolis:2011pv}, where it was built as a coherent state of the Goldstone boson of the vacuum. We would like to conclude by fitting the above construction into the framework of our work.

We begin by selecting a vacuum state $|v_0\ra$ for which $\la v_0 |\Phi|v_0 \ra=v_0$. The classical superfluid configuration $\Phi_0e^{\ri\mu t}$ is parameterized by $\Phi_0$ and $\mu$ which are related by $\mu^2=\lambda (\Phi_0^2-v_0^2)$. A stable solution with real $\mu$ exists for amplitudes $\Phi_0>v_0$. To construct the superfluid state, we can proceed identically to our main analysis, i.e. either in terms of the coherent state of $\Phi$ or of perturbations around the superfluid configuration. Alternatively, we could construct it as a coherent state on top of a vacuum $|v_0\ra$, using the displacement operator of the Goldstone and Higgs bosons. 

The situation somewhat simplifies if we derive the low-energy effective theory of Goldstones by integrating out the Higgs boson. In that case, we only need to utilize the displacement operator of the Goldstone \cite{Nicolis:2011pv}. Our analysis shows that the vanilla coherent state constructed by the field displacement alone acting on the vacuum does not provide a desirable state in the full nonlinear setting. For starters, such a state would be plagued by unrenormalizable divergencies, e.g. infinite energy density. However, even if we were to dismiss this issue, the state would be prone to depletion. To obtain the stable superfluid studied in this work, one would have to supplement the field displacement by the specific squeezing (for one-loop stability) and non-Gaussian modifications (for two-loop stability and beyond). 

Notice that in the full theory, with a propagating Higgs boson, a homogeneous coherent state of merely Goldstones cannot account for the superfluid state. Instead, it creates a state that is confined on the vacuum manifold of the theory. Moreover, degenerate vacua can be understood as coherent states of infinite occupation number of zero-momentum particles with respect to each other \cite{Dvali:2015jxa, Dvali:2015rea, Berezhiani:2024pub}. This is an essential ingredient for constructing topological defects as coherent states \cite{Dvali:2015jxa, Berezhiani:2024pub}.

For the linear sigma model, the tree-level effective theory of Goldstones, to the leading order in derivative expansion, reduces to
\beq
\mathcal{L}=\frac{1}{2}(\partial\pi)^2+\frac{1}{4\lambda v^4}(\partial\pi)^4\,.
\label{outlook:goldstone}
\eeq
In the language of $\pi$, the aforementioned superfluid configuration corresponds to the $\pi=\mu t$ background. This theory is a representative of a broader class of theories described by a generic function of the kinetic term, which are known to provide effective theories of superfluids and have been widely utilized in the cosmology literature (see e.g. \cite{Mukhanov:2005sc}). For the phenomenologically interesting realizations, one is usually interested in backgrounds for which the nonlinear terms dominate the dynamics. Conservatively, such backgrounds are usually deemed to be beyond the regime of validity of the effective theory. However, as we have demonstrated, the mere dominance of nonlinearities is not sufficient to dismiss the background stability and in fact large backgrounds of the form $\pi=\mu t$ are perfectly stable. See e.g. \cite{Creminelli:2019kjy}, for earlier semiclassical discussions of this point, in generic effective theories of the form \eqref{outlook:goldstone}. 

\subsubsection*{Acknowledgments}
We would like to thank Gia Dvali, Justin Khoury and Federico Piazza for discussions. G.C. is supported by the French
government under the France 2030 investment plan, as part of the Initiative d’Excellence
d’Aix-Marseille Université - A*MIDEX (AMX-19-IET-012), and by the  ``action th\'ematique'' Cosmology-Galaxies (ATCG) of the CNRS/INSU PN Astro.

\appendix

\section{Defining the superfluid state without the fluctuation fields}
\label{appstate}
As noted in the introduction, ideally one would like to first quantize the theory of the fundamental field $\Phi(x)$, identify its Hilbert space, and then select the state representing a given configuration, such as the superfluid state above. In the main body of the paper, we introduced fluctuation fields $h(x)$ and $\pi(x)$ and used them to define the superfluid state. Here, we clarify how the stable superfluid state can be defined directly in terms of $\Phi(x)$.

A straightforward way to define a quantum state consists of identifying the fundamental vacuum of the theory $|\Omega\rangle$ and building the state by acting on it with specific operators built out of the field operator and its conjugate momentum. For instance, this procedure is particularly useful when constructing a coherent state in an interacting quantum field theory, as demonstrated in Refs.~\cite{Berezhiani:2020pbv,Berezhiani:2021gph} and summarized in the introduction. In this case, the corresponding operator is simply an exponential of the linear operator in the field and conjugate momentum, generating the field displacement. In contrast, if the exponent is quadratic in the field content, it generates a squeezed state \cite{Berezhiani:2023uwt}.

Equivalently,  the state can be characterized by providing all the correlation functions of the field operator and its conjugate momentum at the time of its definition $t=0$. This is a more convenient approach when the state we build is extremely non-Gaussian and has a non-trivial representation if built using the previous approach. 
The starting point is to determine the one-point functions of the field operator and its conjugate momentum in the chosen state $
|v\rangle$. Here, we impose that these read
\begin{equation}
    \langle v| {\Phi}(\vec{x},0)|v\rangle= \langle v| {\Phi}^\dagger(\vec{x},0)|v\rangle=\frac{v}{\sqrt{2}}, \qquad \langle v| {\Pi}(\vec{x},0)|v\rangle=-\langle v| {\Pi}^\dagger(\vec{x},0)|v\rangle=-{\rm i}{\mu}\frac{v}{\sqrt{2}}\,.
    \label{eq:1pIC}
\end{equation}
To ensure that these initial conditions yield a stationary evolution — so that the time-evolved one-point functions take the form~\eqref{eq:tadpoleC} — the chemical potential $\mu$ must satisfy equation~\eqref{eq:radialU1}. Although this equation is expressed using fluctuation fields, these only appear through their correlation functions, which are ultimately only c-number functions of the parameters $m$, $\lambda$, and $v$. 

The one-point function is insufficient to completely determine the time evolution of the state in the quantum theory, as we miss information about other correlators. We characterize the higher-order correlation functions as if we were acting on the original field $\Phi(x)$ by the linear transformation~\eqref{eq:SFdecomposition}, and then imposing on the fluctuations the same initial conditions that follow from the choice of \eqref{eq:non-gaussianV} as the superfluid state. For example, for quadratic correlation functions, we impose
\begin{flalign}
    &\langle v|  {\Phi}(\vec{x},0)  {\Phi}^\dagger(\vec{y},0) |v \rangle= \frac{v^2}{2}+\frac{1}{2}\left\{\langle \Omega_{h,\pi}|  {h}(\vec{x},0)  {h}(\vec{y},0)|\Omega_{h,\pi}\rangle +\langle \Omega_{h,\pi}|  {\pi}(\vec{x},0)  {\pi}(\vec{y},0)|\Omega_{h,\pi}\rangle\right\}\,,
    \label{eq:quadraticU1} \\
     &\langle v|  {\Phi}(\vec{x},0)  {\Phi}(\vec{y},0) |v \rangle= \frac{1}{2}\Big(v^2+\langle \Omega_{h,\pi}|  {h}(\vec{x},0)  {h}(\vec{y},0)|\Omega_{h,\pi}\rangle -\langle \Omega_{h,\pi}|  {\pi}(\vec{x},0)  {\pi}(\vec{y},0)|\Omega_{h,\pi}\Big)\,.
     \label{eq:quadraticnonU1}
\end{flalign}
Similar conditions apply to the conjugate momenta and to the complex conjugates of the expressions above.\footnote{For the non-$U(1)$ invariant correlation function, we have used the fact that correlators involving an odd number of 
$\pi$ fields vanish at equal times.} The fluctuation fields can be seen as merely auxiliary tools: their correlation functions are to be interpreted as classical c-number functions that encode initial conditions for the correlation functions of $\Phi(x)$. The extension to higher-order correlators then follows straightforwardly.

In the limit of $v\to 0$, the interacting vacuum for fluctuations reduces to the fundamental $U(1)$-invariant vacuum of the theory. This can be explicitly checked at the level of tree-level quadratic correlators, where one finds:
\begin{flalign}
    &\lim_{v\to 0}  \langle v|  {\Phi}(\vec{x},0)  {\Phi}^\dagger(\vec{y},0) |v \rangle =\frac{1}{2}
    \int \frac{\rd^3 k}{2\omega_k (2\pi)^3}\Big(
    \underbrace{1}_{h}
    +
    \underbrace{1}_{\pi}\Big)e^{\ri \vec{k}\cdot (\vec{x}-\vec{y})} =\int \frac{\rd^3k}{ (2\pi)^3 2\omega_k}e^{\ri \vec{k}\cdot (\vec{x}-\vec{y})}\,,\\
     &\lim_{v\to 0}  \langle v|  {\Phi}(\vec{x},0)  {\Phi}(\vec{y},0) |v \rangle =\frac{1}{2}
    \int \frac{\rd^3 k}{2\omega_k (2\pi)^3}\Big(
    \underbrace{1}_{h}
    -
    \underbrace{1}_{\pi}\Big)e^{\ri \vec{k}\cdot (\vec{x}-\vec{y})} =0\,,
\end{flalign}
where $\omega_k^2 = k^2 + m^2$ is the standard vacuum dispersion relation.

In conclusion, with this set of initial conditions, 
the Heisenberg equation for $\Phi(x)$, as given by Eq.~\eqref{eq:EOMSF}, will describe the same 
stationary superfluid configuration previously characterized 
using the fluctuation fields. 
However, the quantization is now carried out 
directly in terms of the fundamental field $\Phi(x)$, as we are never introducing the fluctuations as quantum degrees of freedom, but only as a map to fish out the correct initial conditions for correlators of $\Phi(x)$ and $\Phi^\dagger(x)$.

We conclude this Appendix by discussing how to build the superfluid state explicitly from the vacuum of the theory.
This can be accomplished by inverting the relation~\eqref{eq:SFdecomposition}, and solving for the real and imaginary fields in terms of $\Phi$ and $\Phi^\dagger$.
This has to be supplemented by expressing the relation between the state $|0_{h,\pi}\rangle$ and the free vacuum of the theory, leading to: 
\begin{flalign}
|0_{h,\pi}\rangle=\exp\left\{\ri \hat{f}[\Phi,\Phi^\dagger]\right\}\exp\left\{\ri \hat{S}[\Phi,\Phi^\dagger]\right\}|0\rangle\,.
\label{eq:statePhi}
\end{flalign}

The relation~\eqref{eq:statePhi} is derived by expressing the free vacuum of one set of fields (here, $h(x)$ and $\pi(x)$) as a coherent state built on the free vacuum $|0\rangle$ of another set ($\Phi(x)$ and $\Phi^\dagger(x)$), provided the two are related by a linear transformation. The coherent state is generated by the operators $\hat{f}$ and $\hat{S}$, which are, respectively, the displacement and squeeze operators whose explicit forms are given in Appendix~\ref{App:squeezing}.

\section{The free vacuum for fluctuations as a squeezed coherent state}
\label{App:squeezing}
In this Appendix, we demonstrate that it is possible to represent the free vacuum for fluctuations, $|0_{\pi,h}\rangle$, as a squeezed coherent state built upon the quadratic vacuum of the theory~\eqref{eq:LagrSF}.

To show this result, it is convenient to switch from the complex field basis to the real one
\begin{equation}
 \Phi=\frac{1}{\sqrt{2}}\left(\Phi_1+\ri \Phi_2\right)\,.
\end{equation}
At the fiducial time $t=0$, we introduce the following ladder operator expansion
\begin{equation}
    \Phi_{i}(0,\vec{x})=\int\frac{{\rm d}^3 k}{(2\pi)^3 2\omega_k}\left(a_{i,{k}}e^{\ri \vec{k} \cdot\vec{x}}+a_{i,{k}}^\dagger e^{- \ri \vec{k}\cdot \vec{x}}\right),\qquad \text{with}\quad i=1,2 \,.
\end{equation}
Here, $\omega_k=\sqrt{k^2+m^2}$ is the vacuum dispersion relation. 

Following Refs.~\cite{Berezhiani:2020pbv,Berezhiani:2021gph,Berezhiani:2023uwt}, we introduce a general squeezed coherent state of the form
\begin{equation}
|C\rangle = e^{-\ri f[\Phi_i]} e^{-\ri S[\Phi_i]} |0\rangle\,,
\end{equation}
where $f[\Phi_i]$ is a Hermitian operator linear in the field content, known as the \textit{displacement operator}, and 
$S[\Phi_i]$ is quadratic in the field content, and is known as the \textit{squeezing operator}.
A way to show the equivalence between the $|0_{h,\pi}\rangle$ and the coherent state is to show that there is a choice of ${f}$ and ${S}$ that makes all correlation functions, evaluated at the fiducial time, equal to the analog evaluated over $|0_{h,\pi}\rangle$. This ensures that both states define the same initial conditions and therefore have the same time evolution. To keep the notation compact, here on we assume that all operators are evaluated at the fiducial time $t=0$.

We need to show that the expectation value over the coherent state of a function of the fundamental field variables satisfies
\begin{equation}
    \langle C|\mathcal{O}[\Phi_1,\Phi_2,\Pi_1,\Pi_2]|C\rangle=\langle 0_{h,\pi}|\mathcal{O}[v+h,\pi, \Pi^h, \mu v+\Pi^\pi]| 0_{h,\pi}\rangle\,.
\label{app:exp}
\end{equation}
Here, we introduced the conjugate momenta $\Pi_i=\dot{\Phi}_i$. 
For this proof, we start by parameterizing the operators ${f}$ and ${S}$ as
\begin{flalign}
  &{f} = \int \mathrm{d}^3x \,\left\{ \Pi_i^\text{cl}(\vec{x})\, {\Phi}_i(\vec{x}) - {\Pi}_i(\vec{x})\, \Phi_i^\text{cl}(\vec{x}) \right\}, \\
  &{S} = \int \mathrm{d}^3x \, \mathrm{d}^3y \,
  \begin{pmatrix}
    {\Pi}_i(\vec{x})~,& {\Phi}_i(\vec{x})
  \end{pmatrix}
  \begin{pmatrix}
    \Delta_{11}^{ij}(\vec{x}-\vec{y}) & \Delta_{12}^{ij}(\vec{x}-\vec{y}) \\
    \Delta_{12}^{ij*}(\vec{x}-\vec{y}) & \Delta_{22}^{ij}(\vec{x}-\vec{y})
  \end{pmatrix}
  \begin{pmatrix}
    {\Pi}_j(\vec{y}) \\
    {\Phi}_j(\vec{y})
  \end{pmatrix}\,.
\end{flalign}
Here, we introduced four 
c-number functions to parametrize the displacement operator, and ten functions to parametrize the squeezing operator. Notice that, because we require these operators to be Hermitian, the number of functions required for the squeezed operator is smaller than the naive 16 functions one would expect.
As we show below, these functions specify the initial conditions for the one- and two-point functions of the field operators and their conjugate momenta.

By using the Baker–Campbell–Hausdorff formula, combined with the canonical commutation relations~\cite{Berezhiani:2023uwt}, 
it follows that any correlator, evaluated at the reference time on the coherent state, satisfies
\begin{equation}
    \langle C|\mathcal{O}[\Phi_i,\Pi_i]|C\rangle=\langle 0|\mathcal{O}[\Phi_i^\text{cl}+e^{\ri {S}} {\Phi}_ie^{-\ri {S}},\Pi_i^\text{cl}+e^{\ri {S}} {\Pi}_ie^{-\ri {S}}]|0\rangle\,.
\end{equation}
By comparing the above expression with~\eqref{app:exp}, we fix the functions of the displacement operator to be
\begin{equation}
\Phi^\text{cl}_1=v,\quad \Phi_2^\text{cl}=0,\qquad \Pi^\text{cl}_1=0,\quad \Pi_2^\text{cl}=\mu v\,.
\end{equation}

We are left with fixing the coefficients of ${S}$.
This identification can be provided at the level of the ladder expansion. In other words, we show that there is a choice of the squeezing operator such that the operator $\exp{(\ri S)}\Phi_m \exp{(-\ri S)}$ has the same wavefunction coefficients of the decompositions Eq.~\eqref{eq:condensateDecomposition} and~\eqref{eq:statePhi}. To show this, it is convenient to move to a momentum representation for the squeezed operator according to 
\begin{equation}
 {S}=\int \frac{\rd^3k}{2 (2\pi)^3} \, \sum_{i,j=1,2} \left(\alpha_{ij} a^i_{\vec{k}} a^j_{-\vec{k}}+\text{h.c.}\right)\,.
\end{equation}
Using the Baker-Campbell-Hausdorff formula, we have
\begin{equation}
    e^{\ri S} \Phi_m(\vec{x},0) e^{-\ri S}= \Phi_m+[{\rm i} {S},\Phi_m]+\frac{1}{2}[{\rm i} {S},[{\rm i} {S},\Phi_m]]+\ldots\,\,,
\end{equation}
where ellipses are higher-order commutators.
By explicitly evaluating the first few terms, we find
\begin{flalign}
[{\rm i} {S},\Phi_m]&= {\rm i}\sum_{n=1,2}\int\frac{{\rm d}^3 k}{(2\pi)^3 2\omega_k}\left\{a_{{k}}^n\alpha_{mn}  e^{\ri \vec{k}\cdot \vec{x}}+\text{h.c.}\right\}\,, \\
[{\rm i} {S},[{\rm i} {S},\Phi_m]]&= \sum_{n,l=1,2}\int\frac{{\rm d}^3 k}{(2\pi)^3 2\omega_k}\left\{a_{{k}}^n(\alpha_{l m}\alpha^*_{ln})  e^{\ri \vec{k}\cdot \vec{x}}+\text{h.c.}\right\}\,. 
\end{flalign}
What we learn is that, if we choose $\alpha_{ij}$ in such a way that off-diagonal elements are non-null, we can mix the ladder expansions of $\Phi_{1}$ and $\Phi_2$ using the squeezing operator. Also, each of the commutators does not increase the number of creation and annihilation operators involved in the expansion, which keeps being linear. Therefore, we can schematically write
\begin{flalign}
e^{\ri S}\Phi_me^{-\ri S}=& \sum_{n,l=1,2}\int\frac{{\rm d}^3 k}{(2\pi)^3 2\omega_k}\left\{(\delta_{mn}+\ri \alpha_{ln}\delta_{ lm}+ \alpha_{ln}\alpha^*_{lm}+\dots) a_{{k}}^n e^{\ri \vec{k}\cdot{\vec{x}}}+\text{h.c}\right\}\\
=&\sum_{n=1,2}\int\frac{{\rm d}^3 k}{(2\pi)^3 2}\left\{Z_{mn} a_{{k}}^n e^{\ri \vec{k}\cdot{\vec{x}}}+\text{h.c}\right\}\,.
\end{flalign}
By fixing the coefficients $Z_{mn}$ to match the wavefunction renormalization factors in Eqs.~\eqref{eq:Zpi} and~\eqref{eq:Zh}, it is possible to reconstruct the same correlators obtained using $h$, $\pi$. In particular, we can choose $Z_{11}=\omega^{-1/2}_+Z^h_+$, $Z_{12}=\omega^{-1/2}_-Z^h_-$ and $Z_{21}=\omega^{-1/2}_+Z^\pi_+$,   $Z_{22}=\omega^{-1/2}_-Z^\pi_-$.  For example, using the identifications above, we find that the quadratic correlation functions of $\Phi_2$ read
\begin{equation}
    \langle 0 |e^{\ri S} \Phi_2(\vec{x})\Phi_2(\vec{y}) e^{-\ri S}|0\rangle=\sum_{n} \int \frac{\rd^3 k}{(2\pi)^3 2}|Z_{2n}|^2e^{\ri \vec{k}\cdot(\vec{x}-\vec{y})}=\sum_{a=\pm} \int \frac{\rd^3 k}{(2\pi)^3 2}\left\{\frac{|Z^\pi_a|^2}{\omega^a_k}\right\}e^{\ri \vec{k}\cdot(\vec{x}-\vec{y})}\,,
\end{equation}
matching exactly the equal time version of~\eqref{eq:SSF2}. The same applies to higher-order correlation functions.

The matching between $Z_{mn}$ and the wavefunction coefficients of $h$ and $\pi$ can be performed, for instance, by expressing the squeezing parameters $\alpha_{ij}$ as a perturbative series in the interaction coupling. Since there are four independent wavefunction coefficients, this procedure allows us to fix four linear combinations of the ten. 
Part of the residual six functions are further reduced by matching the ladder decomposition of the conjugate momenta~\eqref{eq:condensateDecomposition2}. The key point to emphasize is that this reconstruction would not be possible without including mixing terms between $a^1$ and $a^2$ in the squeezing operator, which makes the matrix $Z_{mn}$ non-diagonal. Without these, we would miss the fact that $h$ and $\pi$ fail to be described by two independent ladder expansions.

\section{Evaluating cubic correlation functions}
\label{app:correlationfunction}

This Appendix provides details on the procedure adopted to evaluate cubic correlation functions of Section \ref{sec:gaprel}. The procedure is still based on the Dyson series, but has been extended to the case where operators depend on more than one time label. Let us consider two composite operators depending on the space-time coordinates $x=(\vec{x},t)$ and $y=(\vec{y},t')$. Then, the time evolution of their expectation value over the full interacting vacuum is expanded in terms of the free theory according to
\begin{flalign}
  \nonumber  \langle \mathcal{O}(y) ~\mathcal{W}(x)\rangle=&\langle \mathcal{O}_0({y}) ~\mathcal{W}_{0}({x})\rangle  -2\,\text{Im}\int_{-\infty_-}^t \rd^4 z  \, \langle  \mathcal{H}_\text{I}(z) \mathcal{O}_0({y}) \mathcal{W}_0({x})\rangle \\ &-{\rm i}\int_{t}^{t'} \rd^4z\biggr(\langle \mathcal{O}_0({y}) \mathcal{H}_\text{I}(z) \mathcal{W}_0(x)\rangle- \langle \mathcal{H}_\text{I}(z)\mathcal{O}_0(y)\mathcal{W}_0(x)\rangle \biggr)\,.
\label{eq:DysonSeries}
\end{flalign}
Here, expectation values of free fields are taken over the free vacuum $|0_{h,\pi}\rangle$, while the ones involving full Heisenberg fields are evaluated on $|v\rangle$.

Let us consider first the correlation function $\langle \pi(y) \pi(x) h(x)\rangle$, which is the one that enters the one-loop equation of motion for the Goldstone quadratic correlation function. By applying the Dyson series, we find the one-loop correction by evaluating
\begin{flalign}
   \langle \pi(x) \pi(x) h(x)\rangle=&-{\rm i}\int_{t}^{t'}\rd^4 z\left\{ \langle \pi_0(y) \mathcal{H}^{(3)}_I(z) ~\pi_0(x) h_0(x)\rangle- \langle \mathcal{H}^{(3)}_I(z)~\pi_0(y) \pi_0(x) h_0(x)\rangle\right\}\nonumber \\ &-2\text{Im} \int_{-\infty_-}^t \rd^4 z\langle \mathcal{H}^{(3)}_\text{I}(z)~ \pi_0(y) \pi_0(x) h_0(x)\rangle\,,
    \label{cubicAppL2}
\end{flalign}
where the cubic interaction Hamiltonian is determined by~\eqref{eq:cubicSF}.
By performing Wick contractions, we reduce the above expression to
\begin{flalign}
 \langle \pi(x){}& \pi(x) h(x)\rangle = -4\lambda v\, \text{Im} \int_{-\infty}^t \rd^4 z \biggl\{ G_{\pi\pi}(z-y)
\bigl[ 
G_{h\pi}(z-x) G_{\pi h}(z-x) 
+ G_{hh}(z-x) G_{\pi \pi}(z-x) \rangle 
\bigr] \nonumber \\\nonumber
&\quad +G_{h\pi}(z-y) \bigl[ 
G_{\pi h}(z-x)G_{\pi \pi}(z-x) 
+ 3 G_{hh}(z-x)G_{h\pi}(z-x) 
\bigr] \biggr\}  \\&
-2\ri \lambda v \int_t^{t'} \rd t_1 \biggl\{ 
\bigl( G_{\pi \pi}(y-z) - G_{\pi \pi}(z-y) \bigr) 
\bigl[ 
G_{h \pi}(z-x) G_{\pi h}(z-x) 
+ G_{hh}(z-x) G_{\pi\pi}(z-x) 
\bigr] \nonumber \\
&\quad + 
\bigl( G_{\pi h}(y-z) - G_{h\pi}(z-y) \bigr) 
\bigl[ 
G_{\pi h}(z-x)G_{\pi \pi}(z-x) 
+ 3 G_{h h}(z-x) G_{h \pi}(z-x)  
\bigr] 
\biggr\}\,. &
\label{eq:contraction1}
\end{flalign}
Here, $G_{\pi\pi}(x_i-x_j)=\langle \pi_0(x_i)\pi_0(x_j)\rangle$, $G_{\pi h}(x_i-x_j)=\langle h_0(x_i)\pi_0(x_j)\rangle$, $G_{\pi h}(x_i-x_j)=\langle \pi_0(x_i)h_0(x_j)\rangle$ and $G_{hh}(x_i-x_j)=\langle h_0(x_i)h_0(x_j)\rangle$ are the tree-level quadratic correlation functions provided by the expressions~\eqref{eq:SSF2}-\eqref{eq:SSF}.

Next, we perform time integrals and then switch to the momentum-space representation of this correlation function, in terms of the external tri-momentum $q$. Schematically, each contraction takes the form
\begin{equation}
     \langle \pi(y) \pi(x) h(x)\rangle_q\sim \int \rd^3 k_1\rd^3 k_2 \, \delta^{(3)}(q+k_1+k_2) f(q,k_1,k_2;t-t')\,,
     \label{delta}
\end{equation}
where $f$ is the integrand after the time integration is performed.

 Since we are interested in the low-$q$ regime, we drop the $q$ in the delta function and then integrate over one of the internal momenta.
Then, we check the low-$q$ behavior of each contraction. 
Let us denote the terms proportional to $G_{\pi \pi}(z-y)$ and $G_{\pi h}(z-y)$ respectively $A$ and $B$. The first contribution reads
\begin{flalign}
A=-\lambda v\sum_{\alpha=\pm}&\frac{|Z_\alpha^\pi(q)|^2}{(2\pi)^32\omega_\alpha(q)}e^{-\ri \omega_\alpha (t'-t)}\int  \frac{\rd^3 k}{(2\pi)^3}\Bigg(\frac{2|Z_+^h|^2|Z_+^\pi|^2 (\omega_--\omega_+)^2}{\omega_+^2 ((\omega_\alpha(q)+2 \omega_-) (\omega_\alpha(q)+2 \omega_+) (\omega_\alpha(q)+\omega_-+\omega_+)}\nonumber\\ &+\frac{|Z_-^h|^2 |Z_-^\pi|^2}{\omega_-^2 (\omega_\alpha(q)+2 \omega_-)}+\frac{|Z_-^h|^2 |Z_+^\pi|^2+|Z_+^h|^2 |Z_-^\pi|^2}{\omega_-^2 \omega_+^2 (\omega_\alpha(q)+\omega_-+\omega_+)}+\frac{|Z_+^h|^2 |Z_+^\pi|^2}{\omega_+^2 (\omega_\alpha(q)+2 \omega_+)}\Bigg)\,,
\end{flalign}
while the second is determined by the expression
\begin{flalign}
B=-\lambda v\sum_{\alpha=\pm}&\frac{(Z^\pi_\alpha(q))^*Z^h_\alpha(q)}{(2\pi)^3}e^{-\ri \omega_\alpha (t'-t)}\int  \frac{\rd^3 k}{(2\pi)^3\omega_-\omega_+}\sum_{\beta=\pm}\Bigg(\frac{|Z^\pi_\beta|^2-3 |Z^h_\beta|^2 }{(\omega_-(q)+2\omega_-)(\omega_+(q)+2\omega_\beta)}\nonumber\\ & -\frac{|Z^\pi_\beta|^2-3 |Z^h_\beta|^2}{(\omega_+(q)+\omega_-+\omega_+)(\omega_-(q)+\omega_-+\omega_+)}\Bigg)(Z^h_+)^*Z^\pi_+\,.
\end{flalign}
Here, we made implicit the dependence of frequencies and wavefunction coefficients on internal momenta. 
Finally, we expand the integrands in a series of $\omega_+(q)$ and $\omega_-(q)$.
With this procedure, we learn that $B$ approaches a constant in the limit $q\to 0$, and contributes \textit{only} to the constant terms $\sigma_+$ in \eqref{eq:Jq}. 
In particular, at the leading order in the expansion, $B$ approaches the correlation function $\ri \partial_t \langle h(y) \pi(x)\rangle$, and hence provides a loop correction to the kinetic mixing.

The important terms are the ones in $A$. These contain a contribution that diverges as $q^{-1}$, in the small q-limit. Once isolated, it gives rise to $\delta m^2_\pi$ as in \eqref{eq:Beta1}. Subleading constant corrections can be reassembled in the time derivative of the two-point correlation function of $\pi$ and in $\sigma_+$ terms. Therefore, we find
\begin{equation}
   \lim_{{q}/ \mu\rightarrow 0} \, \langle \pi(y) \pi(x) h(x)\rangle_q= (\delta m_\pi^2+\ri \alpha \partial_t) \langle \pi_0(y) \pi_0(x)\rangle_q
    -{ \sigma_+} e^{-{\rm i} \omega_+ (t'-t)}\,.
\label{eq:cubicPiApp}
\end{equation}
 As mentioned, however, the next-to-leading order correction cannot be trusted, as it is of the same order as the terms neglected when setting the momentum $q$ to zero in the delta function of~\eqref{delta}. 
Moreover, let us stress that when the time integral in the first line of~\eqref{cubicAppL2} is evaluated at $t_z = t'$, the integrals over internal momentum cannot be carried out explicitly. However, these contributions are exponentially suppressed in the limit of infinite UV cutoff and can therefore be safely neglected.

Second,  we check correlation functions $\langle 3 h(y) h(x)^2+h(y) \pi(x)^2\rangle$ that enter the equation of motion for the $\langle h(x) h(y)\rangle$ correlator. 
Their expansion after Wick contracting reads:
\begin{flalign}
 &\langle 3 h(y) h(x)^2+h(y) \pi(x)^2\rangle=\\
& -4\lambda v\, \text{Im} \int_{-\infty}^t \rd^4 z \biggl\{ 
G_{hh}(z{-}y) \Bigl[ 9G_{hh}(z{-}x)^2 + 3G_{h\pi}(z{-}x)^2 + 3G_{\pi h}(z{-}x)^2 + G_{h\pi}(z{-}x)^2 \Bigr] \nonumber \\
&\qquad \qquad + G_{\pi h}(z{-}y) \Bigl[ 6G_{\pi h}(z{-}x)G_{hh}(z{-}x) + 2G_{h\pi}(z{-}x)G_{\pi\pi}(z{-}x) \Bigr] \biggr\} \nonumber \\
&\quad - 2\ri \lambda v \int_t^{t'} \rd t_1 \biggl\{ 
\bigl( G_{hh}(y{-}z) - G_{hh}(z{-}y) \bigr) \Bigl[ 9G_{hh}(z{-}x)^2 + 3G_{h\pi}(z{-}x)^2 + 3G_{\pi h}(z{-}x)^2 + G_{h\pi}(z{-}x)^2 \Bigr] \nonumber \\
&\qquad \qquad + \bigl( G_{h\pi}(y{-}z) - G_{\pi h}(z{-}y) \bigr) \Bigl[ 6G_{\pi h}(z{-}x)G_{hh}(z{-}x) + 2G_{h\pi}(z{-}x)G_{\pi\pi}(z{-}x) \Bigr] \biggr\}\,.
\label{eq:contraction2}
\end{flalign}

Among these contractions, only the ones proportional to $G_{hh}(z-y)$ and $G_{hh}(y-z)\rangle$ contain divergent terms, and only when the time-integrals are evaluated on the $t$ extreme. In order to extract them, we can see that these divergences are logarithmic. Therefore, after going in Fourier space, we set the $q$ in the delta over momentum to zero and integrate over one of the internal momenta. Then, we expand all the wavefunction factors and frequencies for big $k$, with $k$ the residual internal momentum, and integrate. This gives us the divergent part described in Eq \eqref{eq:Beta3}. 
The analog correlation function with $h(y)$ replaced by $\pi(x)$ is obtained by a similar procedure. In that case, contractions proportional to $\langle \pi(y) h(z)\rangle$ and $\langle h(z) \pi(y)\rangle$
are the ones containing divergences.

\bibliographystyle{JHEP}
\bibliography{U1coh_final.bib}
\end{document}